\documentclass[useAMS]{mn2e}
\usepackage{natbib_vw,times,epsfig}
\usepackage{amssymb,amsmath}

\bibpunct[, ]{(}{)}{;}{a}{}{,}

\def \aj {AJ}
\def \mnras {MNRAS}
\def \apj {ApJ}
\def \apjs {ApJS}
\def \apjl {ApJL}
\def \aap {A\&A}

\def \pasp {PASP}

\def \hds {Balmer-strong}
\newcommand{\neii}{[Ne\,{\sc ii}]}
\newcommand{\neiii}{[Ne\,{\sc iii}]}
\newcommand{\feii}{Fe\,{\sc ii}}
\newcommand{\mum}{\ensuremath{\mu\mbox{m}}}
\newcommand{\um}{\ensuremath{\mu\mbox{m}}}
\newcommand{\hii}{H\,{\sc ii}}
\newcommand{\ha}{\ensuremath{\mbox{H}\alpha}}
\newcommand{\hb}{\ensuremath{\mbox{H}\beta}}

\def \neiilong {[Ne\,{\sc ii}]\,$\lambda15.5$\um}
\def \neiiilong {[Ne\,{\sc iii}]\,$\lambda12.8$\um}

\def \oiii {[O{\small III}]}
\def \oii {[O{\small II}]}
\def \oiilong {[O{\small II}]\,$\lambda\lambda3727,3730$}
\def \loiii {L[O{\small III}]}
\def \oiiilong {[O{\small III]}\,$\lambda5008$\AA}
\def \oiv {[O{\small IV}]}
\def \loiv {L[O{\small IV}]}
\def \oivlong {[O{\small IV}]\,$\lambda26$\um}
\def \loivlong {L([O{\small IV}]\,$\lambda26$\um)}
\def \nii {[N{\small II}]}

\def \PAHvi {PAH$_{6.2}$}

\title[Nebular emission from dusty galaxies]{Optical vs. infrared
  studies of dusty galaxies and AGN: (I) Nebular emission lines}
\author[V. Wild (et al.)]{
\parbox[t]{\textwidth}{\raggedright 
Vivienne Wild$^1$\thanks{wild@iap.fr},  Brent Groves$^2$, Timothy
Heckman$^3$, Paule Sonnentrucker$^3$, Lee Armus$^4$, David Schiminovich$^5$,
 Benjamin Johnson$^6$, Lucimara Martins$^7$, Stephanie LaMassa$^3$
}
\vspace*{6pt}\\
$^1$ Institut d'Astrophysique de Paris, CNRS, Universit\'e Pierre \& Marie Curie, UMR 7095, 98bis bd Arago, 75014 Paris, France\\
$^2$ Leiden Observatory,  Leiden University, P.O. Box 9513, 2300 RA Leiden, The Netherlands \\
$^3$ Dept. of Physics and Astronomy, Johns Hopkins University, Baltimore, MD 21218, USA \\
$^4$ Caltech, Spitzer Science Center, MS 314-6, Pasadena, CA 91125, USA  \\
$^5$ Dept. of Astronomy, Columbia University, NewYork, NY10027, USA \\
$^6$ Institute of Astronomy, Madingley Road, Cambridge, CB3 0HA, UK\\
$^7$ NAT-Universidade Cruzeirodo Sul, Sao Paulo, Brazil
}

\begin{document}
\maketitle
\begin{abstract}

  Optical nebular emission lines are commonly used to estimate the
  star formation rate of galaxies and the black hole accretion rate of
  their central active nucleus. The accuracy of the conversion from
  line strengths to physical properties depends upon the accuracy to
  which the lines can be corrected for dust attenuation. For studies
  of single galaxies with normal amounts of dust, most dust
  corrections result in the same derived properties within the
  errors. However, for statistical studies of populations of galaxies,
  or for studies of galaxies with higher dust contents such as might
  be found in some classes of ``transition'' galaxies, significant
  uncertainty arises from the dust attenuation correction. In this
  paper we compare the strength of the predominantly unobscured mid-IR
  \neiilong+\neiiilong\ emission lines to the optical \ha\ emission
  lines in four samples of galaxies: (i) ordinary star forming
  galaxies (80 galaxies), (ii) optically selected dusty galaxies (11), (iii) ULIRGs (6), (iv)
  Seyfert 2 galaxies (20). We show that a single dust attenuation curve
  applied to all samples can correct the \ha\ luminosity for dust
  attenuation to a factor better than 2. Similarly, we compare \oiv\
  and \oiii\ luminosities to find that \oiii\ can be corrected to a
  factor better than 3. This shows that the total dust attenuation
  suffered by the AGN narrow line region is not significantly
  different to that suffered by the starforming \hii\ regions in the
  galaxy. We provide explicit dust attenuation corrections, together
  with errors, for \oii, \oiii\ and \ha. The best-fit average
  attenuation curve is slightly greyer than the Milky-Way extinction
  law, indicating either that external galaxies have slightly
  different typical dust properties to the Milky Way, or that there is
  a significant contribution from scattering. Finally, we uncover an
  intriguing correlation between Silicate absorption and Balmer
  decrement, two measures of dust in galaxies which probe entirely
  different regimes in optical depth.
\end{abstract}


\section{Introduction}

In the last few decades large optical spectroscopic surveys have
dramatically improved our knowledge of the global distribution of
galaxy physical properties such as stellar mass, star formation rates,
star formation histories, and both gas phase and stellar
metallicities. Furthermore, at low to intermediate redshifts, the
sheer statistics of current optical spectroscopic surveys allow us to
identify and study significant numbers of rare objects which represent
transient phases in the lifetime of galaxies: starburst,
post-starburst and ``green-valley'' galaxies, major mergers, and
powerful active galactic nuclei (AGN). These transient phases,
although rare, are globally important. For example, black hole growth
occurs primarily in rare, high accretion rate events
\citep{Yu:2002p3731,heckman04}, and it is still debated whether the 
infrequent gas-rich galaxy major mergers
are responsible for the morphological transformation of galaxies
and the build-up of the red sequence 
\citep{Wild:2009p2609,ElicheMoral:2010p5018,Ilbert:2010p5037}. In the
near future, the \emph{James Webb Space Telescope} (JWST) will allow
rest-frame optical spectroscopic observations at high-redshift, where
galaxies are more active in both star formation and black hole growth.

There is an obvious benefit in using the information provided in the
redshift survey to study the properties of the galaxies in detail. For
example, when star formation is the dominant mechanism of excitation,
\ha\ is a very simple and reliable measure of star formation rate in
galaxies. However, it must be accurately corrected for dust
attenuation. In ordinary galaxies, the tight relations between
dust-corrected \ha\ and other indicators of star formation rate from
other wavelength regimes provide circumstantial evidence that dust
attenuation in the optical can be corrected for
accurately. \citet{Kennicutt:2009p4470} suggests that errors on SFRs
caused by the assumed dust curve (attenuation as a function of
wavelength) can be $\pm$10-20\% for typical star-forming galaxies in
the local Universe. However, knowing the ``correct'' dust curve is
crucial. \citet{Muzzin:2009p4926} find that the use of different dust
curves in spectral-energy-distribution (SED) fitting of high redshift
K-selected galaxies can alter parameters such as star formation rates
and star formation timescales by factors of 2-3. The large errors
reported in this study result from the fact that the ``correct''
dust-curve to use at high redshift is largely unconstrained and there
is a large range in the steepness of dust-curves observed along
lines-of-sight to stars in the Magellanic clouds and Milky Way. 

With the complex physical processes that occur during transient phases
in the life-cycle of a galaxy, it is uncertain how far rest-frame
optical studies alone can help us in understanding their true impact.
As dust content increases, and galaxy geometries become more
complicated, corrections for dust attenuation become more
uncertain. How much physical information can we really infer from
optical data alone when faced with extreme star formation, dust
production, AGN activity and complicated spatial geometries of
disturbed galaxies?

In this paper, we compare emission line fluxes in the optical to those
in the mid infra-red where dust attenuation is negligible, except in
small and well known wavelength bands. To test the recovery of \ha\
luminosity from optical data in the presence of significant dust, we
compare and contrast a low-z ``representative'' sample of galaxies
with (i) dusty galaxies selected from the \emph{Sloan Digital Sky
  Survey} (SDSS) based on their Balmer emission line ratios and strong
Balmer absorption lines indicating a possible recent decrease in star
formation; (ii) local \emph{Ultra-luminous Infrared Galaxies} (ULIRGs)
selected from the \emph{Infrared Astronomy Satellite} (IRAS) Bright
Galaxy Sample (BGS) based on their total IR luminosities (L$_{\rm
  TIR}>10^{12}$). The control sample of galaxies arise from the
Spitzer-SDSS-GALEX Spectroscopic Survey (SSGSS) and were chosen to be
``representative'' of the ordinary galaxies within the SDSS and GALEX
surveys. For the AGN in these samples, we further compare the
\oiii/\oiv\ line ratio to a sample of Seyfert 2 galaxies, to test the
recovery of black hole accretion rates from optical \oiii\ emission
line. In a companion paper we will address the effect of dust on the
optical continuum, and thus estimations of star formation histories,
and compare the star formation rates derived from the multi-wavelength
continuum with those from \ha\ luminosity.

Understanding our selected ``unusual'' galaxies is crucial for
understanding galaxy evolution.  The \hds\ galaxies show optical
emission lines indicating obscured AGN, and the correction of their
\oiii\ emission for dust attenuation using different dust-curves
results in them contributing between an insignificant $\sim$1\% and a
much more significant $\sim$10\% to the total \oiii\ luminosity (or
black hole accretion rate) of all narrow-line AGN in the SDSS. This
has important implications for the starburst-AGN connection, for
example, as the strong Balmer absorption lines suggest this accretion
is happening after the starburst has decayed
\citep{Wild:2010p4205}. Low-z ULIRGs are usually associated with major
gas-rich mergers \citep{Sanders:1988p3844}. Some models of galaxy
evolution have used gas-rich major mergers as the driving force to
change the observational properties of the galaxy population over
cosmic time. The mergers lead to powerful starbursts and QSO driven
outflows, resulting in the population of red, quiescent ellipticals
seen in the local universe
\citep[e.g.][]{2006ApJS..163....1H,2006ApJ...636L..81N}. ULIRGs become
increasingly common at earlier epochs, and dominate the star formation
rate density beyond a $z\sim1$
\citep{LeBorgne:2009p4512}. Understanding the basic properties of
ULIRGs at low-redshift, and the limitations of the optical wavelength
regime for identifying and understanding the properties of their power
sources, is crucial in an era when near-infrared (NIR) ground-based
spectrographs and the JWST will soon allow us to study galaxies with
rest-frame optical spectra at increasingly early epochs.

The outline of this paper is as follows. After briefly reviewing the
topic of dust extinction and attenuation in Section \ref{sec:review},
we present the spectral analysis methods and four samples in detail in
Section \ref{sec:samples}.  In Section \ref{sec:results} we compare
the data with a variety of different dust curves commonly used in the
literature. We conclude in Section \ref{sec:disc} by discussing the
implications of our work for measuring star formation rates, black
hole accretion rates, and identifying AGN in optical spectroscopic
surveys. Finally, we provide relations for correcting the \oii, \oiii\
and \ha\ emission lines using a measured \ha/\hb\ flux ratio (Balmer
decrement), assuming a single dust curve which provides a good fit to
all of the samples studied in this paper.

Throughout the paper we quote line luminosities in erg/s and
masses in solar masses. Optical parameters for all samples have been
extracted from the SDSS-DR7, and may differ from those in
earlier releases due to updates in the SDSS reduction pipeline. All
optical wavelengths are quoted as measured in a vacuum, in keeping
with SDSS convention, optical wavelengths are given in \AA\ and mid-IR
wavelengths in \um.

\section{Review of dust attenuation laws}\label{sec:review}

Dust grains in galaxies which intercept and scatter stellar and
nuclear light complicate the analysis of integrated galaxy
spectra. Optical light reaching the observer is effected by dust in
three different ways: (i) absorption; (ii) scattering out of the
line-of-sight; (iii) scattering into the line-of-sight. The first two
combined are usually termed ``extinction'' and all three combined are
termed ``attenuation''. The final ``effective attenuation'' observed
in the integrated light of galaxies results from the integration of
attenuated light over all lines-of-sight to light sources throughout
the entire galaxy\footnote{Throughout this paper we use the term
  ``effective attenuation'' to encompass all effects relating to the
  dust i.e. absorption and scattering, and geometrical effects. We
  note that this differs from some other papers which apply the term
  to ``single-line-of-sight'' attenuation.}. In this section, we
introduce the basics of the effects of dust on studying the integrated
light from galaxies.  \cite{2001PASP..113.1449C}
gives an extensive recent review of the literature, although note that
the terminology differs in places from that adopted here.

The effective attenuation $A_\lambda$, in magnitudes, at a
given wavelength $\lambda$ is given by:
\begin{eqnarray}
A_\lambda &=& -2.5\log \left(\frac{I_{\lambda,obs}}{I_{\lambda,intr}}\right)\\
&=& -2.5\log \left[\exp(-\tau_\lambda)\right]
\end{eqnarray}
where $I_{\lambda,obs}$ is the observed luminosity, $I_{\lambda,intr}$
is the intrinsic luminosity of the source, and $\tau_\lambda$ is the
effective optical depth of the dust. 

Observationally we measure the dimming and reddening of light from
sources, thus deriving an effective attenuation or extinction curve $k_\lambda$:
\begin{eqnarray}
k_\lambda &=& \frac{A_\lambda}{\rm E(B-V)}\\
&=& \frac{\rm E(\lambda-V)}{\rm E(B-V)}+R_V
\end{eqnarray}
where the colour excess is defined by $E(\lambda-V) \equiv
A_\lambda-A_V$ and $R_V$ is the total-to-selective extinction in the
$V$ band.  In the case of extinction, R$_V$ is a function of the grain
composition, size and shape. In the case of effective attenuation
observed in galaxies it is additionally a function of the local
geometry of the dust and light sources and global geometry of the
galaxy (bulge and disk, and inclination). In this work we will use the
term ``dust curve'' to mean either ``effective attenuation'' or
``extinction'' curve, depending on the particular curve under
discussion.

The effective attenuation suffered by the ultra-violet (UV) to optical
stellar continua of starburst galaxies, where the unattenuated stellar
continuum is relatively well understood, has been measured empirically
from the ratio of heavily attenuated to less attenuated spectra
\citep{1994ApJ...429..582C,Calzetti:2000p4473}. The dust-curve is
found to be ``greyer'' (i.e. flatter) than extinction curves measured
along single lines-of-sight towards stars in the Milky
Way\footnote{Commonly used MW extinction curves include
  \citet{Seaton:1979p4492},\citet{1989ApJ...345..245C} and
  \citet{1994ApJ...422..158O}}, as expected from geometrical
effects. Similarly, \citet[][hereafter CF00]{2000ApJ...539..718C} fit
the ratio of far infra-red (FIR) to UV luminosities and UV slope in
starburst galaxies to find an effective attenuation curve that is well
fit by the shallow powerlaw $\tau_\lambda \propto \lambda^{-0.7}$.

Dust curves derived from stellar continua are not necessarily suitable
for the correction of nebular emission lines, however. In starburst
galaxies, nebular emission lines are found to be more attenuated than
the stellar continua \citep[e.g.][]{1994ApJ...429..582C}, indicating
that the light passes through a greater quantity, and possibly
different distribution or composition, of dust. Any difference in the
distribution or composition of dust will result in a different shape
for the dust attenuation curve and therefore errors in the correction
of nebular emission lines using a dust-curve derived from the stellar
continua.  The stronger attenuation of nebular emission lines compared
to the stellar continuum, can be successfully reproduced by models in
which dust is distributed primarily in 2 components: diffuse
interstellar dust, and stellar birthclouds which disperse after a
finite time ($\sim10^7$ years) \citep[][
CF00]{1994ApJ...429..582C,Silva:1998p4065}. This allows the emission
lines, which are ionised by the hottest stars still enveloped in their
dense birthclouds, to experience greater attenuation than the UV
stellar continua, which has an additional contribution from cooler
stars around which the birthclouds have already dispersed.

In Wild et al. (2007) we presented explicitly the form of the
dust-curve for the correction of nebular emission lines in the CF00
dust model:
\begin{equation}\label{eqn:w07}
\frac{\tau_\lambda}{\tau_V} = \frac{A_\lambda}{A_V} =(1-\mu)(\lambda/\lambda_V)^{-n} + \mu(\lambda/\lambda_V)^{-m}.
\end{equation}
We reiterate that we use $\tau_V$ to refer to the final ``effective
attenuation'' suffered by light, including geometrical effects, and
that this model is strictly an ``angle averaged'' model, averaging
over all sight-lines through a galaxy. The first term describes the
slope of the screen-like extinction suffered as the light leaves the
stellar birth-clouds and the exponent is set to match the extinction
observed along lines-of-sight in the MW ($n=1.3$). The second term
describes the additional, much greyer attenuation caused by the
intervening diffuse ISM, where CF00 derive a best-fit exponent of
$m=0.7$. The fraction of the total effective optical depth at 5500\AA\
caused by dust in the diffuse ISM is controlled by the parameter
$\mu$. CF00 showed that the typical value for $\mu$ in their starburst
dataset was about 0.3.  The resulting model attenuation curve, labelled
$\mu_{0.3}$ throughout the paper, naturally lies between the pure
continuum $\tau_\lambda \propto \lambda^{-0.7}$ attenuation
(equivalent to $\mu=1$), and pure screen-like attenuation (equivalent
to $\mu=0$).

Recently, \citet{2008MNRAS.388.1595D} applied and extended the two
component CF00 dust model to a sample of local galaxies with
multi-wavelength photometry and optical \ha\ and \hb\ emission line
measurements, finding an excellent match to the data.

We note that an additional complication arises in the case of nebular
emission from the narrow line region (NLR) of an AGN, where a
different dust geometry could lead to a different effective
attenuation curve than that found for nebular emission from \hii\
regions. Additionally, for galaxies with contributions to nebular
emission from both star formation and an AGN NLR, the use of the
Balmer decrement for the correction of NLR emission lines may be
somewhat inappropriate.  In general it is reasonably assumed that the
Hydrogen recombination lines are dominated by the emission from \hii\
regions in all but the strongest AGN. However, this is not the
case for higher ionization and higher excitation lines where emission
from the NLR can dominate. Thus, while the Balmer decrement will show
how much dust is in \hii\ regions, there is no reason to believe that
light from the NLR would pass through the same amount of
dust\footnote{The difference in the intrinsic Balmer decrements
  between NLR and \hii\ regions (due to different gas temperatures),
  will be small relative to the difference caused by
  attenuation.}. Therefore the extinction correction applied to these
other lines, such as [\oiii]$\lambda5007$\AA, may be incorrect. We
investigate these effects in Section \ref{sec:bhar}.

\subsection{Attenuation of nebular emission lines}\label{sec:nir}

In order to determine the amount of attenuation suffered by optical
nebular emission lines, ideally one would observe a pair of lines
whose intrinsic ratio is independent of physical conditions and where
one of the pair lies at a wavelength long enough that attenuation is
negligible \citep[see also discussion
in][]{Osterbrock:2006p4475}. Hydrogen recombination line ratios that
include one line in the NIR wavelength range have been shown to be a
powerful tool for constraining the dust attenuation curves for nebular
emission lines in galaxies
\citep{Calzetti:1996p4710,2001PASP..113.1449C}. The data are well fit
by a screen-like MW extinction curve, which has since been favoured
for the correction of emission lines for dust attenuation. These
results suggest that a substantial fraction of the attenuation of
emission lines must arise from dust in a foreground screen-like
geometry (e.g. the birthclouds in the CF00 model). However, the
samples studied to date have been small, containing only local
starburst galaxies. Extending this study to a larger sample of
galaxies, including more ordinary starforming galaxies and dustier
galaxies is the purpose of this paper.

Sample sizes using ground based NIR observations are restricted
primarily by the difficulty resulting from contamination of
observations by OH sky lines. \emph{Hubble Space Telescope}
observations of NIR Paschen-$\alpha$ line have been used to reveal the
detailed distribution of star formation in galaxies relatively
unimpeded by dust \citep{Boker:1999p4704}. The detection of dust lanes
in some galaxies in this sample indicate some level of extinction
still exists even at 1.875\mum. Uncertainties in the form of
dust-curve in the NIR regime therefore cause a small additional source
of uncertainty when using NIR emission lines to constrain dust-curves
in the optical.

With the launch of the \emph{Spitzer Space Telescope}, with the
Infrared Spectrograph \citep[IRS,][]{Houck:2004p2735} on board, high
quality, moderate resolution mid-IR spectra are available for the
first time for a large number of low-$z$ galaxies. Extinction in the
mid-IR is negligible, except in small wavelength regions centered on
9.7 and 18\mum\ where strong absorption can arise from the stretching
and bending modes of amorphous silicate grains.  While H recombination
lines are found in the Mid-IR, they unfortunately tend to be too weak
to be easily measured.  Two of the strongest and easily observable
emission lines in the mid-IR arise from neon; \neii\ 12.8\mum\ and
\neiii\ 15.5\mum. In typical \hii\ regions, over 95\% of neon is in
these two states (Ne$^+$ and Ne$^{2+}$), thus we expect that the sum
of these two lines will trace very well the ionized gas, and thus \ha,
in galaxies. Indeed, \citet{Ho:2007p4951} have already suggested that
these mid-IR Neon lines could provide a useful new indicator of star
formation rate in galaxies. They have shown that \neii+\neiii\
luminosity correlates strongly with both total infrared luminosity in
galaxies and the NIR Br$\alpha$ emission line in giant \hii\
regions. \citet{DiazSantos:2010p5114} have shown that \neii\
luminosity correlates strongly with that of Pa$\alpha$ in star-forming
regions in LIRGS. Neon has the added benefit of being an abundant
element, especially as it is not depleted onto dust grains.  Therefore
in this work we use the sum of these two lines as a tracer of the
\emph{unextinguished} emission from ionized gas in these
galaxies. Although these lines lie on the edge of the silicate
absorption features, in general the extinction caused by this is
minimal as we verify in Section \ref{sec:samples}.  Combining Spitzer
spectra with optical SDSS spectra provides a new and extremely
powerful dataset, in which both optical and mid-IR emission lines have
been measured much more accurately than was possible in the past.

Starburst/\hii\ region modelling
\citep{2001ApJ...556..121K,Levesque:2010p4655} suggests that for
typical twice Solar metallicity \hii\ regions, the ratio
L(\neii+\neiii)/L(\ha) should be between 0.5-1.0, demonstrating how
luminous the mid-IR Ne lines are. The range is caused by ionization
effects, arising from different stellar models, the mean age of the
ionizing stellar population, and ionizing flux densities. There is one
major disadvantage of using the [Ne]-to-\ha\ ratio to constrain the
optical dust-curve in galaxies, in that the intrinsic ratio is
directly dependent upon the metallicity of the gas, greatly expanding
the range of possible values. For Solar metallicity \hii\ regions the
models predict L(\neii+\neiii)/L(\ha) should be smaller by a factor of
0.6.  However, we expect that this is not a dominant concern in the
samples studied in this paper as we are exploring massive, dusty, and
therefore metal-rich, galaxies.  We discuss the impact of variations
in the intrinsic ratio where relevant. We refer the reader to
\citet{Ho:2007p4951} for further details on the formation of the neon
lines in \hii\ regions.

In AGN the ionizing spectrum is much harder than found in star-forming
regions, meaning that neon starts to populate higher ionization
states. The sum of the mid-IR neon lines still traces the ionized gas
very well, but the ionisation is offset to higher values due to the
harder radiation field. In addition, the harder radiation field of AGN
also create partially (or X-ray) ionized regions, which have a higher
L([\neii])/L(\ha) ratio. AGN photoionization models suggest
L(\neii+\neiii)/L(\ha) $\sim 2$, a factor of 2-4 higher than \hii\
regions \citep{Groves:2004p4703}\footnote{These line ratio predictions
  can be obtained using the publicly available ITERA software package
  \citep{Groves:2010p4895}}. Again, we will indicate the effect of
this result where relevant.

\section{The Samples}\label{sec:samples}

\begin{figure*}
\centering
\includegraphics[scale=0.6]{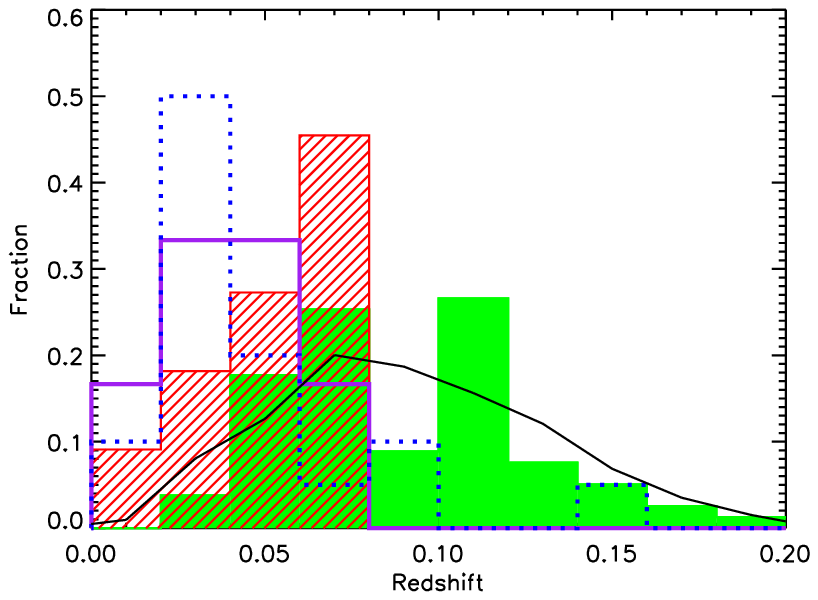}
\includegraphics[scale=0.6]{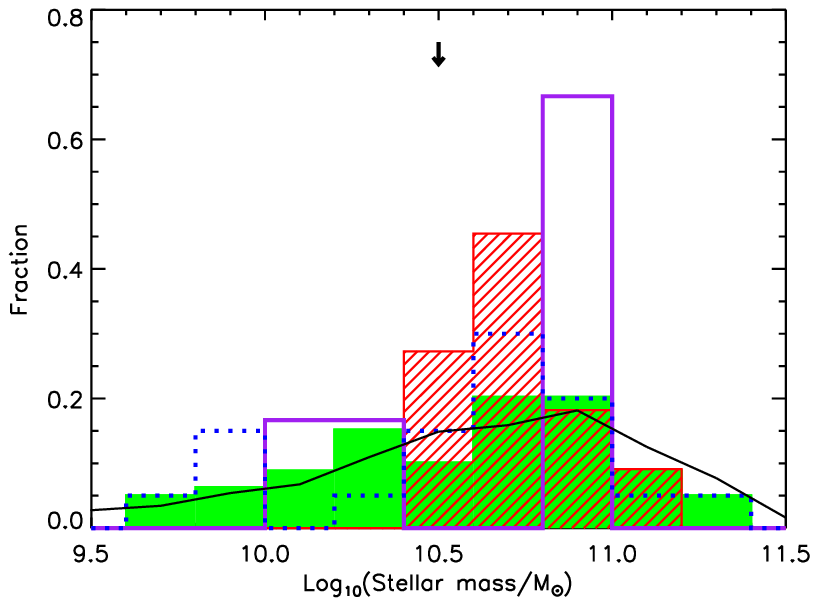}
\includegraphics[scale=0.6]{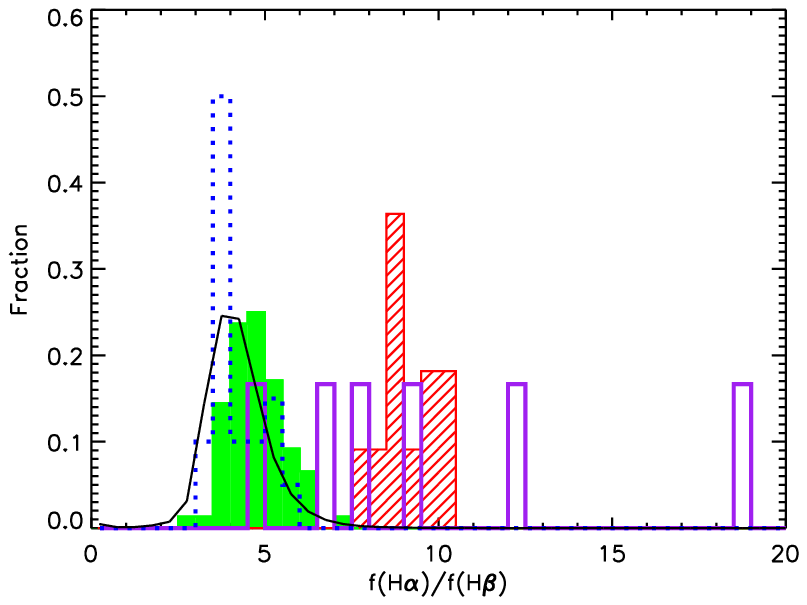}
\caption{The distributions of redshifts (left), stellar masses
  measured from SDSS 5-band optical photometry (center) and \ha/\hb\
  ratios (right) of the 4 galaxy samples: SSGSS (green, shaded), \hds\
  (red, line-filled), ULIRG (purple), Seyfert (blue, dotted).  The
  black line shows the distributions for SDSS-DR7 galaxies which have
  been spectroscopically observed and have a spectral SNR$_{\rm g}>8$.
  The downward arrow in the central panel shows the position of the
  observed bimodality in the galaxy mass distribution (because the
  histograms are not corrected for sample incompleteness the
  bimodality is not noticeable in this figure). }\label{fig:z}
\end{figure*}

In this section we give the methodology used to measure the nebular
emission lines, followed by the selection criteria of our four samples
of galaxies. All samples have SDSS optical 5-band imaging and fibre
spectroscopy, together with Spitzer IRS spectroscopy. The SDSS
specobjid, SDSS-MPA rowindex, RA and Dec of the \hds, ULIRG and
Seyfert samples are given in Tables \ref{tab:psb}, \ref{tab:ul} and
\ref{tab:sey} to allow easy identification of the objects in the
catalogues used in this paper. The equivalent information for the
SSGSS sample is available online at
http://www.astro.columbia.edu/ssgss. For a quick visualisation of the
samples: Figure \ref{fig:z} presents their distributions in redshift,
stellar mass and Balmer decrement (\ha/\hb\ flux ratio), Figure
\ref{fig:bpt} shows their positions on the BPT diagram
\citep[\nii/\ha\ vs. \oiii/\hb\ emission line flux
ratios][]{1981PASP...93....5B}, and in Figure \ref{fig:spoon} we place
all our samples on the \citet{2007ApJ...654L..49S} diagram of \PAHvi\
equivalent width vs. the strength of the 9.7\um\ silicate
absorption. We discuss these figures further in Section
\ref{sec:comparison}.

The raw optical emission-line measurements for all SDSS-DR7 galaxies
are provided on the SDSS-MPA web
pages\footnote{http://www.mpa-garching.mpg.de/SDSS/DR7/} and details
are given in \citet{2003MNRAS.346.1055K} and
\citet{2004MNRAS.351.1151B}; errors have been rescaled according to
the information provided on the web site. Note that all broad emission
line objects have been identified by the SDSS automated pipeline and
are not included within the SDSS-MPA sample. Briefly, spectral
synthesis models [Bruzual \& Charlot (2003); 2008 version using the
MILES spectral library, S{\'a}nchez-Bl{\'a}zquez
et~al. (2006)]\nocite{2003MNRAS.344.1000B,miles} are used to fit the
stellar continuum, including stellar absorption features, and emission
lines are measured from the residual of the data and model
fit. Therefore \emph{underlying stellar absorption is accurately
  accounted for when measuring optical emission line
  fluxes}. Intervening Galactic extinction is corrected for prior to
the calculation of the emission line fluxes. The quality of the fits
obtained is excellent.

For all samples, IRS spectroscopy was carried out in staring mode and
1D spectra were extracted using the standard Spitzer IRS Custom
Extraction (SPICE). For the \hds\ galaxies the ``optimal extraction''
option was chosen \citep{Narron:2007p4861}, which weights the
extraction by the object profile and the signal-to-noise of each pixel
and is suitable for faint sources. Individual 1D spectra were combined
using custom algorithms.

To maximise consistency between the heterogeneous samples in this
paper, the final low-resolution 1D spectra of all samples were
reanalysed using the same custom algorithms and analysis packages
where appropriate. To obtain \neiilong\ and \neiiilong\ fluxes for the
SSGSS and Seyfert samples we used the mid-IR spectral fitting package
PAHfit \citep[v1.2][]{Smith:2007p949} which fits Gaussian profiles
with widths set according to the instrument module. PAHfit was run in
the rest-frame using a screen-like dust model, and although dust
attenuation is tiny at the wavelengths of the Ne lines, for complete
consistency between samples we removed the correction applied by the
code \citep[Eqn. 4,][]{Smith:2007p949}. Our final results are
unchanged whether the minimal dust correction is applied or not. Where
high resolution spectroscopy is available, we use the high-resolution
measurements of \neii\ and \neiii\ due to the smaller error bars. For
the \hds\ sample, the \neii\ and \neiii\ emission line fluxes were
measured interactively from the short-high (SH) IRS spectra using the
2-step baseline and Gaussian line-fit module from the SMART IDEA
package
\citep{Higdon:2004p2644}\footnote{http://isc.astro.cornell.edu/SmartDoc/LineFitting}.
For the ULIRG sample, fluxes were taken from high resolution
spectroscopy, given in Table 4 of \citet{Armus:2007p964}. To ensure
consistency between the four samples, where possible\footnote{ We found
that PAHfit does not fit the continuum well for ULIRGs with strong
Silicate absorption, and therefore the emission line measurements were
unreliable in these cases. } we verified that
\neii\ and \neiii\ fluxes measured with PAHfit or directly from the
low-resolution spectra for the \hds\ and ULIRG samples were equal
within the errors to those measured from the high-resolution spectra
(with and without the minimal correction for attenuation applied in
PAHfit).  For the Seyfert sample, \neiilong\ fluxes
were found to match within the errors those measured in the
high-resolution mode, given in Table 6 of \citet{2010arXiv1007.0900L}.

\oivlong\ was measured from high-resolution (LH) IRS spectroscopy in
all cases (measurement from the low-resolution spectra is not possible
due to a nearby FeII line). For the \hds\ sample, \oiv\ line fluxes
were measured interactively using the SMART IDEA package as described
above. For the ULIRG and Seyfert samples \oiv\ line fluxes were taken
from the tables in Armus et al. (2007) and \citet{2010arXiv1007.0900L}
respectively.

One of the major components of a star-forming galaxy spectrum in the
mid-IR is the broad emission features generally accepted to result
from the vibrational modes of polycyclic aromatic hydrocarbons (PAHs).
While we do not study the PAH features in detail in this paper, PAHfit
outputs their fluxes and equivalent widths which we use to provide
further clarification of the nature of our objects where necessary.

To visualise the different dust and AGN properties of our different
samples in Section \ref{sec:comparison}, we measure the equivalent
width of \PAHvi\ and the strength of the 9.7\um\ silicate absorption.
We attempted in both cases to follow closely the methods as described
in \citet{2007ApJ...654L..49S}, and we used the ULIRG sample to verify
that our results are consistent with those given in
\citet{2007ApJ...654L..49S} over the full range of the measurements.

\subsection{The SSGSS sample}
Our control sample is the Spitzer-SDSS-GALEX Spectroscopic Survey
\citep[SSGSS,][]{ODowd:2009p4395,Johnson:2009p2742}, a Spitzer Legacy
survey (P.I. D.~Schiminovich) which targeted 100 ``ordinary'' galaxies
from the SDSS spectroscopic survey and GALEX surveys for follow-up
mid-IR spectroscopy with the Infrared Spectrograph
\citep[IRS,][]{Houck:2004p2735} using the Spitzer Space Telescope. The
galaxies are magnitude limited to the SDSS spectroscopic magnitude
limit of $r<17.7$ and additionally were selected to have GALEX-NUV
fluxes $>1.5m$Jy for the faint sample (40 galaxies) and $>5m$Jy for
the bright sample (60 galaxies). They were chosen such that their
mass, colour, star formation rate and redshift distribution are
representative of galaxies in the SDSS survey. A description of the
survey, together with data tables containing SDSS and mid-IR
measurements and derived parameters, are available at the team
website\footnote{http://www.astro.columbia.edu/ssgss/}. In this paper
we make use of the IRS low resolution spectroscopy of 80 SSGSS
galaxies. The IRS spectra of the remaining galaxies are either
incomplete due to problems during observations or reductions, or
suffer from considerable noise at short wavelengths, 5 of these caused
by a solar flare. A further 4 galaxies with median per-pixel-SNR$<10$
in their SDSS spectra are removed, as the spectra were of insufficient
SNR to allow accurate stellar continuum fitting around the Balmer
emission lines.  Further details of the SSGSS can be found in
\citet{ODowd:2009p4395} and \citet{Johnson:2009p2742}.

\subsection{The dusty Balmer-strong sample}
Our first sample of galaxies with significant dust contents are 12
\hds\ galaxies selected from the SDSS-DR4 to have an observed \ha/\hb\
flux ratios greater than 8 (where a ratio close to 3 indicates little
dust) and optical line ratios that indicated the possible presence of
an AGN. Additionally, they were selected to have strong Balmer
absorption lines, which might indicate a recent sharp decline in star
formation rate (``post-starburst''). However, the strong Balmer
absorption lines might alternatively be caused by unusual dust
properties rather than star formation history\citep[e.g. as postulated
in][]{2000ApJ...529..157P}. This is the subject of a future
paper. Both high- and low-resolution IRS spectra were obtained for all
of these 12 galaxies during Spitzer Cycle-4 (P.I. T.~Heckman).
Further observational details of this sample are provided in Appendix
\ref{app:psb}. One of the 12 galaxies was subsequently discovered to
have double peaked narrow \ha\ and \nii\ lines, together with
significant underlying broad emission which was not accounted for in
the original emission line measurements. Its IRS spectrum is dominated
by emission from the AGN. Attempts to robustly remeasure the \hb\ line
accounting for the double peak were unsuccessful due to significant
stellar absorption, and its Ne/\ha\ and PAH/\ha\ ratios suggests that
it is not in fact unusually dusty. We therefore remove it from all
further discussion, leaving 11 \hds\ galaxies.

\subsection{The ULIRGS}
Our second sample of dusty galaxies is composed of the 10 ULIRGs ($\rm
L_{IR}>10^{12}L_\odot$) from the IRAS Bright Galaxy Sample
\citep{Soifer:1987p2581}; the Spitzer IRS low- and high-resolution
observations of these galaxies have been presented in detail in
\citet{Armus:2007p964}. Of the 10 BGS ULIRGs, 6 are found in the SDSS
spectroscopic sample [IRAS15327+2340 (Arp220), IRAS 08572+3915, IRAS
15250+3609,IRAS 09320+6134 (UGC 5101),IRAS 12112+0305], 3 do not lie
within the SDSS footprint (IRAS 05189-2524, IRAS 22491-1808, IRAS
14348-1445), and 1 is optically too faint to have been targeted [IRAS
12540+5708 (Mrk 231) has an SDSS $r$-band magnitude of 19.4]. All 6 of
the ULIRGs with SDSS spectra are late stage mergers
\citep{Veilleux:2002p4924}.

\subsection{The Seyfert sample}
In Section \ref{sec:bhar} we compare the high ionisation emission
lines in the dusty galaxies with those of a sample of relatively 
dust-free (low Balmer decrement) Seyfert galaxies. This is composed of the top 20
\oiii$\lambda5007$ flux emitters out of all SDSS-DR4 main-sample
galaxies classified as Seyfert galaxies based on their position on the
BPT diagram.  They have \oiii\ fluxes greater than
$4\times10^{-14}$erg/s/cm$^2$, corresponding to a luminosity of
$\sim10^{41}$erg/s at the median redshift of the sample.  Both high-
and low-resolution IRS spectra were obtained, with high
signal-to-noise ratio. Further details of the Seyfert sample, along
with a study of their X-ray properties can be found in
\citet{LaMassa:2009p5102} and \citet{2010arXiv1007.0900L}.

\subsection{Sample comparison}\label{sec:comparison}

\begin{figure}
\hspace*{-0.5cm}
\includegraphics[scale=0.5]{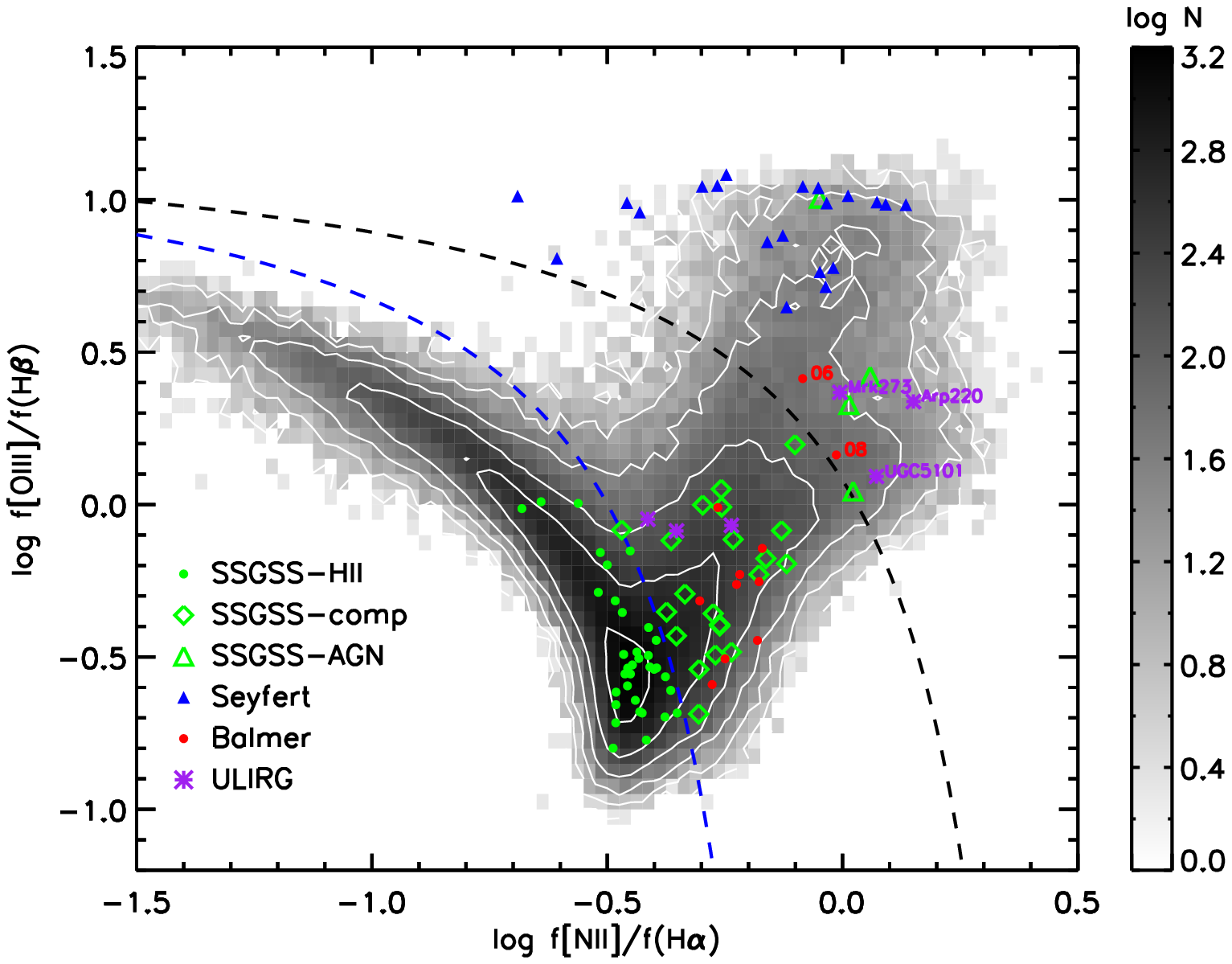}
\caption{The BPT diagram \citep{1981PASP...93....5B}: the narrow
  emission line flux ratios BPT$_x=$f\nii$_{\lambda6585}$/f(\ha)$_{\lambda6565}$
  and  BPT$_y=$f\oiii$_{\lambda5008}$/f(\hb)$_{\lambda4863}$ for the SDSS-DR7
  spectroscopic galaxy sample (gray scale), SSGSS starforming
  galaxies (green dots, 35/35), SSGSS AGN (green open triangles, 4/4),
  SSGSS composite galaxies (green open diamonds, 21/21), \hds\ galaxies
  (red dots, 11/11), ULIRGs (purple stars, 6/6), Seyferts (blue
  triangles, 20/20). Only galaxies with all 4 lines detected at $>3\sigma$
  confidence are plotted. The dashed lines indicate our division
  between starforming/composite (blue) and composite/AGN (black)
  samples.  These demarcation lines define the three optical classes
  (starforming, composite, AGN). To aid the discussion, the \hds\
  galaxies and ULIRGs in the AGN region have been
  labelled.}\label{fig:bpt}
\end{figure}

\begin{figure}
\hspace*{-0.5cm}
\includegraphics[scale=0.7]{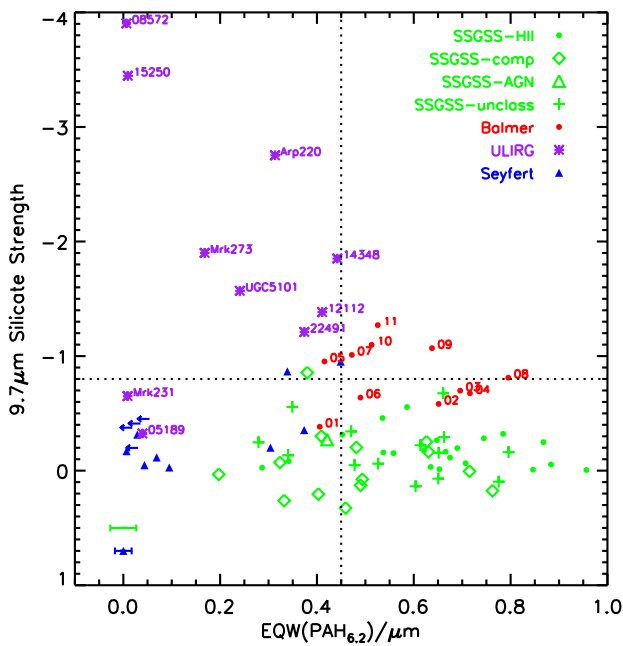}
\caption{The strength of the 9.7\um\ silicate absorption vs. the
  equivalent width of the 6.2\um\ PAH feature, following the
  measurement methods presented in \citet{2007ApJ...654L..49S}. Note
  that the x-axis is linear in order to better distinguish our
  samples.  The dotted lines are two of the dividing lines given by
  Spoon et al. (2007) to separate galaxies with no significant
  Silicate absorption (horizontal) and ordinary star-forming/starburst
  galaxies (vertical). Median errors are shown in the bottom left for
  the Seyfert and SSGSS samples, median errors are smaller than the
  symbol size for the \hds\ and ULIRG samples. Only those galaxies
  with mean per-pixel-SNR$>2$ between 5.4 and 5.9\um\ are
  plotted. Symbols are: SSGSS starforming galaxies (green dots,
  25/35), SSGSS AGN (green open triangles, 2/4), SSGSS composite
  galaxies (green open diamonds, 14/21), SSGSS galaxies with
  insufficient SNR in one or more optical emission lines to provide an
  optical BPT classification (green crosses, 20/20), \hds\ galaxies
  (red dots, 11/11, labelled), ULIRGs (purple stars, 6/6, labelled),
  Seyferts (blue triangles, and upper limits marked by blue arrows,
  13/20). }\label{fig:spoon}
\end{figure}

The samples that we will compare and contrast in this paper were
targeted for different purposes and thus any conclusions drawn must
take account of potential systematics caused by their different
selections.

The redshift distribution of the 4 samples is shown in the left panel
of Figure \ref{fig:z}. The black line shows the redshift distribution
of the SDSS spectroscopic galaxy catalog with $r$-band extinction
corrected petrosian magnitude\footnote{Petrosian magnitudes and radii
  are estimates of total magnitudes suitable for all types of
  galaxies. They are derived by the SDSS photometry pipelines, with
  details found here:
  http://www.sdss.org/dr7/algorithms/photometry.html} in the range
$14.5<r<17.77$.
The Seyfert and ULIRG sample are flux limited in nebular emission and
IR flux respectively, and lie predominantly at low redshifts. The
SSGSS galaxies were selected to be representative of the full SDSS
sample, and therefore their redshift distribution tracks closely that
of the full SDSS survey. The \hds\ sample is limited to $z<0.07$ to
ensure that the stellar population sampled by the optical fibre
through which the spectrum is taken lies within the central region of
the galaxy close to the AGN.

The stellar mass distribution of the 4 samples is shown in the central
panel of Figure \ref{fig:z}. The stellar masses\footnote{Courtesy
  of Jarle Brinchmann, publicly available at
  http://www.mpa-garching.mpg.de/SDSS/DR7/ } have been estimated from
stellar population model fits to the optical 5-band SDSS photometry,
using a Bayesian analysis similar to that described in
\citet{2003MNRAS.341...33K} and \citet{2005ApJ...619L..39S}. Most
samples cover the full range in stellar mass probed by the SDSS,
although the \hds\ galaxies do not cover the low-mass end of the
distribution. Note that the histograms are \emph{not} corrected for
incompleteness effects due to sample or survey flux limits, and
therefore are not suitable for drawing conclusions about the underlying
distributions of galaxy mass.  To aid the reader, the position of the
galaxy mass bimodality is marked by a downward arrow. The right hand
panel of Figure \ref{fig:z} shows the Balmer decrement distributions
of the 4 samples, showing how the \hds\ and ULIRG samples are
significantly dustier on average than the SSGSS and Seyfert samples. 

In Figure \ref{fig:bpt} we show the position of the four samples on
the BPT diagram, with the underlying grayscale showing the
distribution of spectroscopically observed galaxies in the SDSS. Only
those galaxies with SNR$>3$ in all four emission lines are
plotted. This is the most commonly used optical diagram for diagnosing
the presence of an AGN. Throughout this paper, we distinguish 4
classes of galaxies: (i) \emph{AGN-dominated}, with emission-line
ratios which place them above the stringent theoretical starburst
criterion of \citet{2001ApJ...556..121K}, or have \nii/\ha$>0.0$; (ii)
\emph{star-formation dominated}, which lie below the observationally
determined demarcation line of \citet{2003MNRAS.346.1055K}, or have
\nii/\ha$<-0.5$; (iii) \emph{composite} objects, which lie between the
AGN-dominated and star-forming dominated samples; we term them
``composite'' to indicate that the emission may be caused by a
combination of both AGN and starformation \citep[][although see
caution below]{2003MNRAS.346.1055K}; and (iv) \emph{unclassified}
objects with non-existent or weak (SNR$<3\sigma$) emission lines which
do not allow classification into the first three samples. 

We note that the BPT diagram does not provide a firm demarcation
between the samples and that the demarcation lines are not motivated
by robust theoretical predictions; we use the lines only to provide a
convenient broad classification. \citet{Levesque:2010p4655} present
up-to-date modelling of the ``maximum starburst'' demarcation line,
placing it much closer to the observed star forming branch in the SDSS
than the \citet{2001ApJ...556..121K} line. Shocks, high metallicity,
and ionisation by old stars can also cause galaxies to lie in the
``composite'' class \citep{Stasinska:2008p4785}. We note that most of
the \hds\ galaxies have detectable molecular H$_2$ lines, as do the
majority of ULIRGs \citep{Higdon:2006p4736}. In the case of ULIRGs
these lines are thought to arise primarily from photodissociation
regions (PDR) associated with star formation.  The H$_2$/PAH flux
ratio of the \hds\ galaxies is exactly that expected for star forming
galaxies (Ogle et al. in prep.), indicating that shocks are similarly
not a dominant excitation mechanism in the \hds\ galaxies. This
suggests that the high \nii/\ha\ line ratios in these samples are in
general caused by a central AGN, rather than shocks. Signal-to-noise
ratio limits placed on emission lines may preferentially cause very
high-metallicity galaxies and LINERs caused by shocks and/or
ionisation by old stars to be ``unclassified'' in the SSGSS sample
\citep{CidFernandes:2010p4744}. However, none of these caveats affect
the results of this paper which do not rely on a precise
identification of weak AGN, or on the precise positioning of pure AGN
or starforming galaxies on the BPT diagram. Where possible we include
unclassified SSGSS galaxies in the figures.

Out of the 80 SSGSS galaxies, 4 are AGN-dominated, 21 are composite
objects, 35 are star-forming and 20 are unclassified. Out of the 11
\hds\ galaxies 2 are AGN-dominated and the remainder composite
objects.  3 ULIRGs are AGN-dominated (Arp220, Mrk273 and UGC5101) and
3 are composite objects (08572, 12112, 15250).  All Seyferts clearly
lie in the AGN dominated region by definition. Figure \ref{fig:bpt}
supports our assumption that the intrinsic L(\neii+\neiii)/L(\ha)
ratio will not vary greatly between our samples. Most galaxies in the
samples lie in the solar and greater region of this line ratio
diagnostic diagram \citep{Groves:2006p4862,Levesque:2010p4655}.

Finally, in Figure \ref{fig:spoon} we follow
\citet{2007ApJ...654L..49S} and plot the equivalent width (EQW) of the
6.2\um\ PAH feature against the strength of the 9.7\um\ silicate
absorption. We plot only those objects with mean SNR-per-pixel$>2$ in
the wavelength regions used to define the continuum for the
measurement of the silicate absorption at 9.7\um\ and with total
flux/error$>3$ in the 6.2\um\ PAH feature. The lower branch is
interpreted as a mixing between AGN and starformation activity, and
the upper branch as an evolution from a dust obscured, to dust
unobscured nuclear starburst or AGN (see Rowan-Robinson \& Efstathiou
2009).  Because of its focus on both dust and AGN activity,
Fig. \ref{fig:spoon} is particularly useful for characterising our 4
different samples:
\begin{description}
\item [{\bf Galaxies in the SSGSS}] occupy the bottom right corner of
  the diagram as expected for ``ordinary'' star-forming galaxies or
  weak narrow-line AGN. They have strong EQW(\PAHvi) and weak silicate
  absorption. The composite galaxies in this sample have a lower mean
  EQW(\PAHvi) than the star-forming galaxies, strengthening the
  argument that these galaxies contain an AGN. The large scatter, and
  significant number of objects with apparent silicate emission, is
  indicative of the generally lower quality of these spectra compared
  to our other samples. SSGSS galaxies with optical emission lines
  which are too weak to classify them on the optical BPT diagram do
  not differ in mid-IR properties from their classified counterparts.
\item [{\bf The \hds\ galaxies}] generally have stronger Silicate
  absorption than SSGSS galaxies, placing them between the ULIRG and
  SSGSS samples and therefore, within the common interpretaion of the
  diagram, at the end of their evolution from a dusty starburst. M82
  is the prototype galaxy for this region of the diagram.
\item [{\bf The ULIRGS}] are a mixed group, as has been discussed at
  length in the literature. Although our sample is small, it is
  interesting to note that the ULIRGs with particularly strong
  silicate dust absorption (08572, 15250), show optical emission line
  ratios indicative of a mixture of star formation and AGN and/or
  shocks. Three out of the four ULIRGs with weaker silicate absorption
  (Arp220, Mrk273, UGC5101), have AGN-dominated optical emission
  (i.e. implying little contribution from star formation, except
  possibly in the case of extreme shocks). This is consistent with the interpretation of the upper
  branch as an evolution from dusty starburst galaxies in the top-left
  to galaxies undergoing less intense, less dusty star formation in
  the bottom-right, with dust-obscured AGN activity occurring
  throughout the evolution.
\item [{\bf The Seyferts}] are generally found to have very weak PAH
  equivalent widths and little Silicate absorption, and therefore lie
  in the region of the diagram commonly interpreted as being occupied
  by strong AGN which heat the dust to high temperatures. 2/20 show
  some silicate absorption together with stronger EQW(\PAHvi). The
  median (mean) BPT$_y$ values of the 4 Seyferts with
  EQW(\PAHvi)$>0.2$ are 0.78 (0.75), compared to 1.01 (0.99) for the
  remainder of the sample. This suggests that emission from star
  formation contributes to the optical emission lines even in the
  Seyfert branch of the BPT diagram.  These results are consistent
  with the interpretation of the lower branch of Fig. \ref{fig:spoon}
  as mixing between AGN and starformation activity.
\end{description}

\subsection{Aperture Corrections}\label{sec:aper}

``Aperture bias'' describes the effects caused by the fact that a
fixed aperture or slit width may probe only a small fraction of a
galaxy's total light, and this fraction is a function of galaxy radius
and redshift. The physical size probed by the SDSS 3\arcsec\ fibre
varies from under 1kpc for the majority of the ULIRG sample to around
10kpc at the maximum redshift of the SSGSS galaxies. Fortunately the
results in this paper rely upon line ratios or equivalent widths
rather than absolute luminosities, and therefore aperture bias in this
usual sense is less relevant.

However, comparing line strengths from the IRS spectra with those from
the SDSS spectra leads to other aperture related issues. The finite
(and different) slit and fibre apertures may mean that the IRS and
SDSS spectra sample a different fraction of the same galaxy, and this
will cause measured line ratios to differ from the true values. There
are two important issues: (i) light is lost from the slit/fibre due to
the PSF; (ii) light is lost from the outer regions of extended
sources, depending on the size of the aperture and the accurate
positioning of the aperture on the source. In this paper the precise flux
calibration between the two datasets is relatively unimportant, but it
is still important that we minimise any additional scatter in the
observed relations. To help the reader to understand the limitations
of our data, we outline here the flux calibration procedures of the
mid-IR and optical spectra.

The IRS PSF increases substantially from the blue to the red
wavelength extremes of the instrument, and therefore each IRS slit has
a different width designed so as not to loose too much of the light at
the reddest ends\footnote{The short-low (SL) IRS module covering the
  wavelength range 5.2-14.5\um\ has a slit aperture of
  3.6\arcsec$\times$57\arcsec.  The long-low (LL) IRS module covering
  the wavelength range 14.0-38.0\um\ has a slit aperture of
  10.6\arcsec$\times$168\arcsec. At high resolution, the short-high
  (SH) module covers 9.9-19.6\um\ with a slit aperture of
  4.7\arcsec$\times$11.3\arcsec\, and the long-high (LH) module covers
  18.7-37.2 with a slit aperture of
  11.1\arcsec$\times$22.3\arcsec.}. The spectral extraction procedure
accounts for the varying PSF along the slit by normalising the flux at
each wavelength according to the fraction of light missed from a
calibration stellar point-source. 

The SDSS DR7 spectra have been spectrophotometrically calibrated using
the PSF magnitudes of stars observed on the same plate as the galaxies
\citep{AdelmanMcCarthy:2008p4857}.  To aid aperture corrections, the
SDSS-MPA emission line measurements have been recalibrated to the
3\arcsec\ fibre magnitudes. Therefore, to match the stellar
PSF-normalised mid-IR spectra, we have removed this recalibration,
returning the flux level of the SDSS emission lines to the stellar
PSF-normalised values.

The use of stellar PSF spectrophotometric calibration for both optical
and mid-IR spectra allows good relative calibration for point
sources, but inaccuracies will still arise in the case of extended
sources. For centrally concentrated sources, such as circumnuclear
starbursts in ULIRGs or AGN NLR emission, the point-source
approximation is a good one. However, for extended sources, the
measured mid-IR to optical line ratios may depend on the relative size
of the apertures, and the relative surface brightness distribution of
the sources. Both effects are difficult to correct for accurately,
especially in the case of radial differences within the sources (for
example, in AGN vs. starforming nebular emission). We have therefore
not attempted any further aperture correction, and will caution the
reader at points in the text where this effect may be relevant.

\section{Results}\label{sec:results}

\begin{figure*}
\includegraphics[scale=0.6]{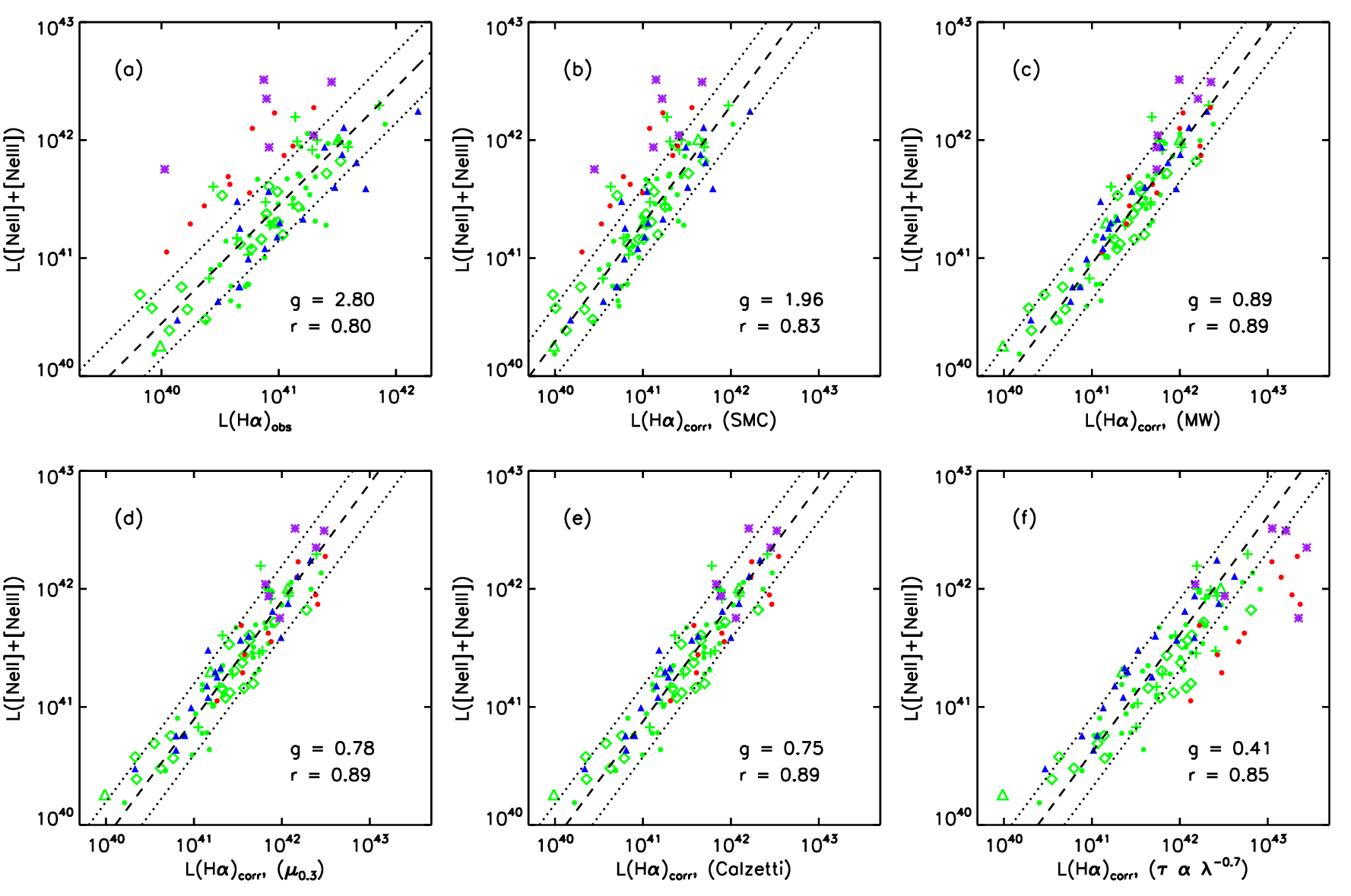}
\caption{The optical \ha$_{\lambda6565}$ vs. mid-IR
  \neiilong+\neiiilong\ luminosities in erg/s. {\it Panel (a):} with
  no correction of \ha\ for dust attenuation (note the different
  x-axis range in this panel). {\it Panels (b-f):} after correcting
  \ha\ for dust attenuation using different dust-curves as described
  in the text; note the offset x-axis scales to panel (a).  The panels
  are ordered by increasing ``greyness'' of the dust-curve used, from
  the screen-like SMC and MW extinction-curves to the attenuation
  curves derived from the continua of starburst galaxies (Calzetti and
  $\tau\propto\lambda^{-0.7}$). See text for references to the
  particular dust-curves. The dashed line shows a maximum-likelihood
  line fit to the \emph{SSGSS-starforming galaxies only}, with a
  factor of two above and below the best-fit relation indicated by
  dotted lines.  The best fit slope to the \emph{SSGSS-starforming
    sample only} ($g$) is indicated in the bottom right of each panel,
  together with the Pearson correlation coefficient ($r$) for the
  \emph{SSGSS-starforming sample only}. The statistical errors on $g$
  are a few percent. Only those galaxies with Ne, \ha\ and \hb\ lines
  detected at better than 3$\sigma$ are plotted.  Symbols are: SSGSS
  starforming galaxies (green dots, 35/35), SSGSS AGN (green open
  triangles, 3/4), SSGSS composite galaxies (green open diamonds,
  19/21), \hds\ galaxies (red dots, 11/11), ULIRGs (purple stars,
  6/6), Seyferts (blue triangles, 19/20). }\label{fig:neha}
\end{figure*}

\begin{figure*}
\includegraphics[scale=0.6]{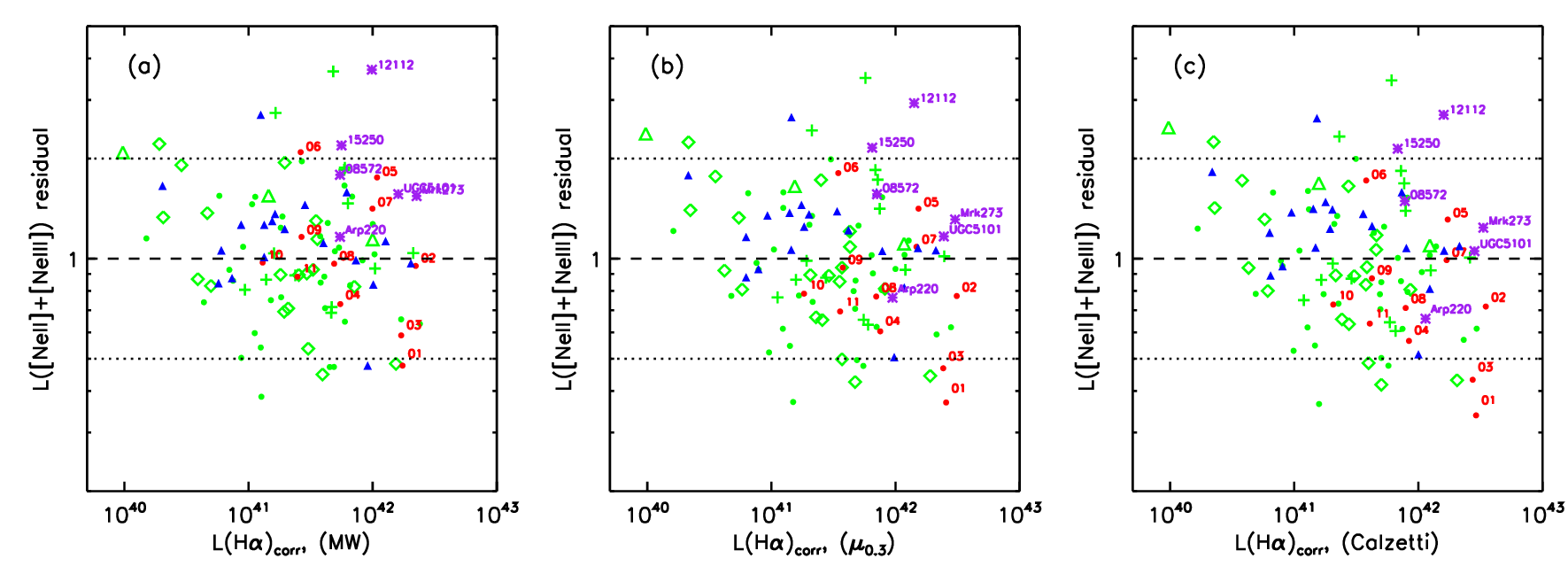}
\caption{The residual [Ne] luminosity from the best-fit linear
  relation between dust corrected \ha\ and [Ne] luminosities, as a
  function of \ha\ luminosity in erg/s. From left to right, \ha\ has
  been corrected for dust attenuation using the MW, $\mu_{0.3}$ and
  Calzetti dust-curves. The offset of the AGN-dominated sources from
  the starformation-dominated sources is clear. The mean offset for
  the Seyfert sample is a factor of 1.26 for the $\mu_{0.3}$ dust
  curve. Symbols and lines as in Fig. \ref{fig:neha}. }\label{fig:neha_resid}
\end{figure*}

In Figure \ref{fig:neha} we plot \ha\ vs.  \neiilong$+$\neiiilong\
(from now on [Ne]) luminosities for all of the samples, firstly with
\ha\ uncorrected for dust attenuation, and then corrected with five
different dust-curves selected from the literature to cover the full
range in possible greyness. The panels are ordered by increasing
greyness of the dust-curve used, from the screen-like
extinction-curves (SMC and MW), to galaxy continuum derived
attenuation-curves (Calzetti and $\lambda^{-0.7}$). The SMC curve was
evaluated from the tabulated results of \citet{Pei:1992p4812} with an
R$_V=3.1$; the MW curve was taken from \citet{1994ApJ...422..158O}
with an R$_V=3.1$; the $\mu_{0.3}$ curve is from \citet{wild_psb}; the
Calzetti curve is from \citet{Calzetti:2000p4473} with R$_V=4.05$; the
$\lambda^{-0.7}$ curve is from CF00 as discussed in Section
\ref{sec:review}.

In each panel a maximum-likelihood best-fit linear relation is fit to
the SSGSS-starforming sample allowing for errors in both $x$ and $y$
axis quantities and fixing the intercept to be zero (dashed line).
Twice and half the best-fit [Ne] luminosity for each \ha\ luminosity
is indicated as dotted lines to aid comparison between the panels of
the scatter between the two quantities. The best-fit gradient ($g$,
equal to the best-fit intrinsic [Ne]/\ha\ ratio) and Pearson's
correlation coefficient ($r$) are given in the bottom right of each
panel, both evaluated using the SSGSS-starforming sample only.

In panel \emph{a} it is immediately evident that without correcting \ha\
for dust attenuation the scatter between \ha\ and [Ne] emission
luminosities is large, even for the SSGSS galaxies with normal
levels of dust. For the dusty \hds\ and ULIRG samples, there is a
clear excess of [Ne] for their observed \ha\ luminosity as expected
for dusty galaxies. 

As we move through the dust laws, from least to most grey [panels
\emph{b} to \emph{f}], we see steady trends. The best-fit intrinsic
[Ne]/\ha\ ratio monotonically decreases, and the correlation
coefficient is maximal for the MW, $\mu_{0.3}$ and Calzetti
dust-curves. Focussing on the dusty \hds\ and ULIRG samples, we notice
that these move from lying above the relation defined by the
SSGSS-starforming galaxies in panels \emph{a} and \emph{b}, to lying
below the relation in panel \emph{f}. This is caused by their under-
and over-correction for dust attenuation using the different
dust-curves.

In Figure \ref{fig:neha_resid} we show the samples more clearly by
plotting the residual of the [Ne] luminosity from the linear fit using
the MW, $\mu_{0.3}$ and Calzetti dust curves. The mean and variance of
each sample in these figures is informative for comparing the
different dust curves. For the dusty ULIRGs the mean is 2.0, 1.6, 1.5
for the MW, $\mu_{0.3}$ and Calzetti dust-curves respectively
i.e. they are significantly offset for all dust-curves (see below for
further discussion). For the \hds\ sample, the mean is 1.1, 0.9 and
0.8, suggesting perhaps that a dust curve in between the MW and
$\mu_{0.3}$ is most appropriate but any offset between this extreme
dusty sample and ordinary starforming galaxies is very small.  For the
SSGSS starforming sample, the mean is 1 by definition for all
dust-curves. The variance is $\sim$0.15 for all dust-curves, i.e. the
scatter is less than a factor of 2 and there is no strong trend with
\ha\ luminosity for the starformation dominated galaxies.

From these figures we can identify the expected offset of the AGN
caused by a higher intrinsic [Ne]/\ha\ ratio. For the Seyfert sample
and $\mu_{0.3}$ dust curve the mean offset is 1.26, i.e.  about a 25\%
enhancement in [Ne] for their \ha\ luminosity. The magnitude of this
offset does not correlate with position on the BPT diagram or PAH
equivalent width, although the size of the sample is too small to rule
out the possibility that those with a smaller offset have a greater
contribution to star formation in their lines. Averaged over the
samples, the offset is seen only for the Seyfert galaxies, and the
three SSGSS-AGN which lie in or close to the Seyfert region on the BPT
diagram. The SSGSS composite sample have a mean of 1 for all
dust-curves, which suggests that the AGN in general does not
contribute significantly to the Ne/\ha\ line ratio in these composite
galaxies.  However, the trend upwards at low \ha\ luminosities is
driven by composite objects, suggesting there may be a small effect at
low star formation rates.  The observed average enhancement in the
[Ne]/\ha\ ratio of the ULIRGs noted above, could be due to small
number statistics, the dust curve (i.e. a shallower dust curve would
be more appropriate), or the presence of deeply buried AGN. We note
that the ULIRGs are only offset by at most a factor of 3, i.e. dust
corrected \ha\ luminosity tracks mid-IR [Ne] luminosity to better than
a factor of 3 even in extreme dusty and disturbed objects.  We will
return to these points below.

The maximum likelihood fits provide a measure of the dust-free ratio
of [Ne] to \ha\ luminosity in our IRS/SDSS datasets, when star
formation is the dominant excitation mechanism for the line emission:
\begin{equation}
{\rm L[Ne]} = g\times{\rm L(H\alpha)_{corr}}
\end{equation}
where L[Ne] = L(\neiilong)+L(\neiiilong) and $g=0.89, 0.78, 0.75$ for
the MW, $\mu_{0.3}$ and Calzetti dust-curves respectively. While it is
encouraging that the measured intrinsic ratio lies exactly at the
center of the range predicted by the models for twice Solar
metallicity \hii\ regions (see Section \ref{sec:nir}), a precise
comparison between models and data would only be possible after much
greater care has been taken to perform precise aperture
corrections. In the case of the AGN, the emission line source is
point-like and therefore the relative calibration between the mid-IR
and optical spectra should be the most accurate. The fact that the
average [Ne]/\ha\ ratio is a little lower than predicted may be due to
contamination by star formation in a subset of the Seyfert sample (see
Section \ref{sec:comparison}).

In the following subsection we study the differences between the dust
curves in more detail.

\subsection{Comparison of common dust attenuation laws}

\begin{figure*}
\centering
\includegraphics[scale=0.7]{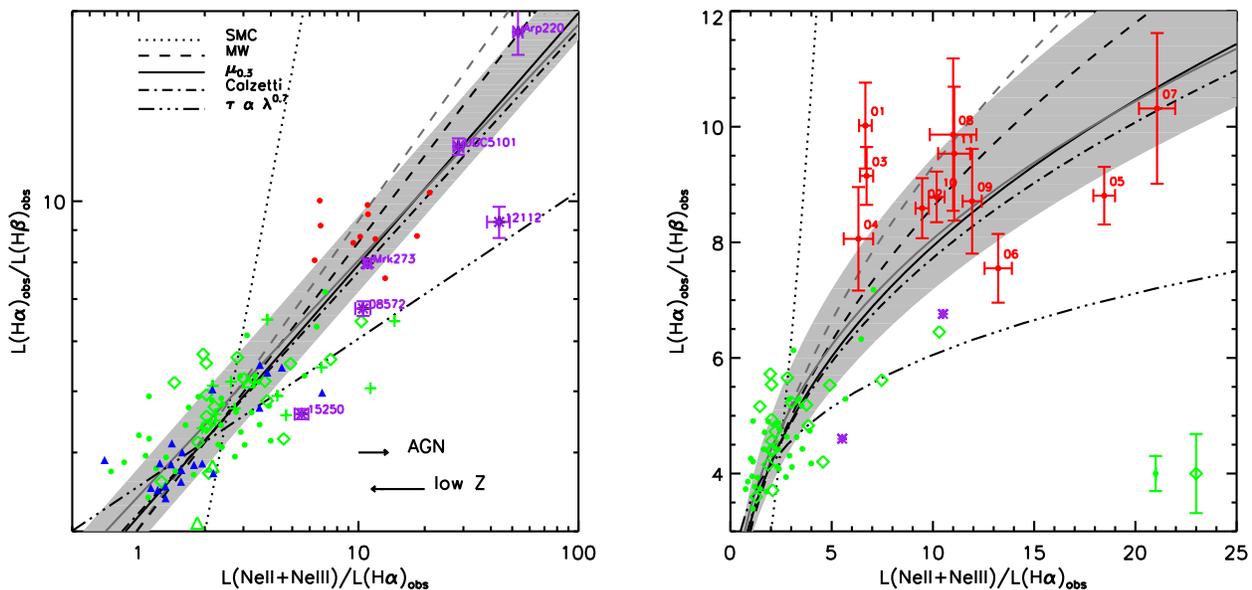}
\caption{The observed [Ne]/\ha\ and \ha/\hb\ flux ratios. \emph{Left:}
  A logarithmic scale allows all samples to appear, and all
  starforming, AGN, composite and unclassified objects are
  plotted. \emph{Right:} A linear version of the same plot, with the
  axes restricted to focus on the less extreme galaxies. Only
  star-forming and composite galaxies are plotted. Overplotted as
  black lines are the predicted ratios from different dust curves: the
  SMC extinction curve with an R$_V=3.1$ \citep{Pei:1992p4812}; a MW
  extinction curve with R$_V=3.1$ \citep{1994ApJ...422..158O}; the
  $\mu_{0.3}$ attenuation curve for emission lines \citep{wild_psb};
  the Calzetti attenuation curve for starburst continua with
  R$_V=4.05$ \citep{Calzetti:2000p4473}; the $\lambda^{-0.7}$
  attenuation curve for starburst continua from CF00. The grey lines
  additionally mark a different MW extinction curve with R$_V=3.2$
  \citep[dashed grey,][]{Seaton:1979p4492,Nandy:1975p4502} and our new
  best-fit dust-curve from Section \ref{sec:fitdust} (solid grey) with
  $\mu=0.4$. In all cases the intrinsic Ne/\ha\ ratio is taken from
  the best-fit to the SSGSS-starforming galaxies. The grey shaded
  region shows the theoretical range in observed [Ne]/\ha\ ratio for a
  twice solar stellar population and the $\mu_{0.3}$ dust-curve, and
  the bottom arrow in the left hand panel indicates the theoretical
  decrease in this ratio as metallicity decreases from twice Solar to
  Solar. The upper arrow in the left hand panel indicates the average
  observed increase in [Ne]/\ha\ ratio where an AGN is the dominant
  ionisation source (i.e. a factor of 1.26 observed for the Seyfert
  galaxies).  For clarity, error bars are indicated for the ULIRGs in
  the left panel, and for the \hds\ galaxies in the right
  panel. Median errors are indicated in the right panel for the SSGSS
  starforming and composite galaxies. Symbols as in
  Fig. \ref{fig:neha}. }\label{fig:dustlaws}
\end{figure*}

In Fig. \ref{fig:dustlaws} we plot the observed [Ne]/\ha\ vs. \ha/\hb\
luminosity ratios for all of our samples. In the left-hand panel we
use a logarithmic scale to allow the dustiest ULIRGs to appear. In the
right-hand panel we focus on the less extreme objects, and only
include galaxies where their optical emission line ratios indicate
starformation is a significant emission line source (classified as
composite or starforming in the optical BPT diagram,
Fig. \ref{fig:bpt}). Overplotted as black lines are the predicted line
ratios from the different dust curves presented in the previous
subsection. In each case the intrinsic [Ne]/\ha\ ratio is set from the
fit to the SSGSS-starforming galaxies as described above.

Although there are several factors which can move galaxies within this
diagram, dust has the strongest effect, increasing both the Balmer
decrement and observed [Ne]/\ha\ ratio by an order of magnitude over
the range probed by our samples.  The different dust-curves predict
different relations between these two line ratios, even when we allow
the intrinsic [Ne]/\ha\ ratio to be a free parameter fit by the
data. As shown in the previous section, the MW, $\mu_{0.3}$ and
Calzetti dust-curves provide the best fits to the SSGSS-starforming
galaxies. Figure \ref{fig:dustlaws} shows that the $\mu_{0.3}$ and
Calzetti attenuation curves also pass through the center of the \hds\
sample. 

The scatter of the SSGSS-starforming galaxies about the best-fit
dust-curves is large, and this may have several origins. Firstly, the
errors on the line flux measurements, particularly in \hb\ are
significant. Secondly, we expect some variation due to different
metallicities. The grey shaded region shows the range predicted by the
theoretical models for twice Solar \hii\ regions, and the arrow in the
left hand panel shows how this region moves for Solar \hii\
regions. The 3 galaxies with the lowest measured [Ne]/\ha\ ratio lie
significantly to the left of the $2Z_\odot$ prediction, but within the
$Z_\odot$ prediction. These 3 galaxies also have some of the highest
\oiii/\hb\ and \neiii/\neii\ ratios of the sample, indicative of
harder ionisation perhaps resulting from lower metallicity star
formation. The single Seyfert galaxy with a [Ne]/\ha\ ratio less than
1 has the lowest \nii/\ha\ ratio of the whole sample, unusually low
for Seyfert galaxies, and likely associated with accretion of
low-metallicity gas \citep{Groves:2006p4862}. 

While low metallicities can explain the scatter of SSGSS-starforming
galaxies to the left of the dust-curve predictions, the scatter to the
right is unlikely to be caused by metallicity variations as scatter in
this direction requires significant and unlikely neon abundance
increases (i.e.~greater than twice solar).  This leads us to the
third, and most interesting, cause for scatter in the line ratio
relation: intrinsic differences in the type or distribution of dust in
galaxies\footnote{Face-on galaxy orientations can also cause increased
  scattering into the line-of-sight and thus greyer dust curves
  \citep{Rocha:2008p4515}. }. In a 2-component dust model, increasing
the fraction of dust in the diffuse ISM relative to the dense
birthclouds increases the contribution from scattering and makes the
dust-curve greyer. In the CF00 emission line dust model given in
Eqn. \ref{eqn:w07} this relative amount of diffuse vs. dense dust is
controlled by the parameter $\mu$. By increasing $\mu$ to 0.9
(i.e. 90\% of the effective optical depth from the dust comes from the
ISM, and 10\% from the stellar birthclouds), we can describe the lower
envelope of the SSGSS galaxies. While this qualitative result is
suggestive, a further detailed study including orientation,
metallicity and ionisation effects would be required to confirm that
these galaxies do indeed have a significantly different dust geometry.

It is evident that the 3 ULIRGs classified as composite galaxies by
their optical line ratios also appear to favour a greyer dust curve
than the dusty \hds\ galaxies. However, as we discuss further below,
it is also plausible that the high Ne/\ha\ ratio observed in these
ULIRGs, as well as in some SSGSS-composite galaxies, is simply due to
a deeply buried AGN.

We summarise the results up to this point by noting that a single
appropriate dust curve, combined with an accurately measured Balmer
decrement, can recover the intrinsic optical emission line
luminosities to better than a factor of two in the majority of
ordinary starforming and dusty galaxies. Our data shows that this
remains true for galaxies with dust contents up to an $A_V$ of
4.4. However, while the use of a MW, $\mu_{0.3}$ or Calzetti
dust-curves for large samples of galaxies will cause the least overall
bias in the final dust-corrected emission line luminosities, the
effect of a greyer dust-curve in individual galaxies should not be
understated. A galaxy with a very typical measured \ha/\hb\ ratio of
4.5 would have its \ha\ luminosity, and thus star-formation rate,
corrected by a factor of around 2.5 using the average dust curve,
whereas a factor of 7 would be more appropriate if the majority of its
dust were in the diffuse ISM.

\subsection{Fitting the dust curve}\label{sec:fitdust}

Throughout the remainder of the paper we will focus on results using
the CF00 $\mu$ dust-curve as it allows greater flexibility than the
traditional dust-curves in the calculation of errors, and the
discussion of variations in dust-curves within the sample as presented
in the previous section. Additionally, it is the only dust-curve
specifically motivated and justified for the correction of nebular
emission lines. The MW dust-curve assumes a screen-like dust
distribution, and the Calzetti dust-curve is intended for correcting
the continuum light from galaxies.  Rather than assuming $\mu=0.3$ as
derived by CF00, we can fit the [Ne] vs. \ha\ luminosities of the
SSGSS-starforming galaxies to obtain the best-fit $\mu$ and intrinsic
[Ne]/\ha\ ratio together with errors. We use a maximum-likelihood
linear fit, accounting for errors on both quantities, and fix the
intercept to be zero\footnote{Allowing the intercept to be non-zero
  results in the same fitted parameters within the errors.}. Formal,
statistical errors on the derived $\mu$ are at the level of 5\%. The
errors obtained from bootstrap resampling with replacement of the data
are more significant, and we quote these errors below, and use them to
obtain errors on the dust correction formulae presented in Section
\ref{sec:sfr} and \ref{sec:bolcorrn}.

We find $\mu=0.4\pm0.2$, i.e. on average 40\% of the optical depth at
5500\AA\ arises in the ISM of the galaxies. This is consistent within
the errors with the $\mu=0.3$ derived by CF00 for their sample of
local starburst galaxies. The intrinsic [Ne]/\ha\ ratio fit at the
same time is $g=0.6\pm0.1$. The resulting curve is shown as the solid
gray line in Figure \ref{fig:dustlaws} and given explicitly as:
\begin{equation}\label{eqn:w10a}
  \frac{\tau_\lambda}{\tau_V} = \frac{A_\lambda}{A_V} = 0.6(\lambda/5500)^{-1.3} + 0.4(\lambda/5500)^{-0.7}
\end{equation}
where $\lambda$ is in \AA. Given a measured \ha/\hb\ flux ratio, the
optical depth at any wavelength can be obtained from:
\begin{equation}\label{eqn:w10b}
  \tau_\lambda  = 3.22\ln\left(\frac{2.86}{f(H\alpha)/f(H\beta)} \right) \frac{\tau_\lambda}{\tau_V}
\end{equation}
where $\tau_\lambda/\tau_V$ is given in Eqn. \ref{eqn:w10a}
$A_\lambda= 1.086\tau_\lambda$ is the attenuation in magnitudes, and a
Case B recombination ratio of 2.86 is assumed.

\subsection{Dense vs. diffuse dust}

\begin{figure}
\centering
\includegraphics[scale=0.4]{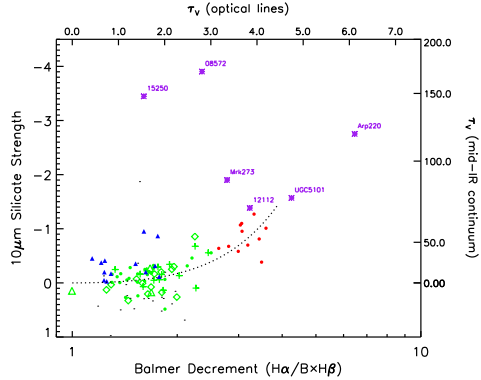}
\caption{The optical Balmer decrement vs. Silicate 9.7\um\ absorption
  strength.  The Balmer decrement is defined with an intrinsic ratio
  of B=2.86 (3.1) for galaxies classed as starforming/composite (AGN)
  from their optical emission lines. Thus, a Balmer decrement of 1
  indicates no optical dust attenuation. The top axis indicates the
  corresponding optical $\tau_V$, assuming the best-fit $\mu_{0.4}$
  dust-curve derived in Section \ref{sec:fitdust}. The right-hand axis
  indicates the corresponding mid-IR $\tau_V$ from the zero-age
  starburst models of Rowan-Robinson \& Efstathiou (2009).
  Overplotted as a dotted line is the best-fit relation to the
  SSGSS-starforming, composite and \hds\ galaxies as described in the
  text. Only those galaxies with \ha\ and \hb\ detected at greater
  than 3$\sigma$ and mean per-pixel-SNR$>2$ between 5.4 and 5.9\um\
  are plotted. Symbols are: SSGSS starforming galaxies (green dots,
  25/35), SSGSS AGN (green open triangles, 3/4), SSGSS composite
  galaxies (green open diamonds, 14/21), \hds\ galaxies (red dots,
  11/11), ULIRGs (purple stars, 6/6), Seyferts (blue triangles,
  13/20).  }\label{fig:dust}
\end{figure}

Extinction in the mid-IR continuum is dominated by the stretching and
bending modes of amorphous silicate grains with two strong features at
9.7\mum\ and 18\mum. Importantly for this section, it is evident from
the many theoretical models that \emph{any} significant silicate
absorption, as found in the ULIRGs and many of the \hds\ galaxies,
implies optical extinction at levels of $\tau_V \sim 50-100$
\citep[e.g.][]{Sirocky:2008p4863,2009MNRAS.399..615R}. It is therefore
impossible for ionising UV continuum and associated line radiation to
pass through these clouds directly.

In Figure \ref{fig:dust} we plot the Balmer decrement
[\ha/(B$\times$\hb) where B is the unattenuated ratio] against the
9.7\um\ Silicate absorption strength for all four samples. For \hii\
regions we set $B = 2.86$ for Case B recombination at 10,000K and
electron density of $10^2-10^4$/cm$^{-3}$
\citep{Osterbrock:2006p4475}. For AGN we set $B=3.1$
\citep{2006MNRAS.372..961K}.  Surprisingly, we see a correlation
between these two measures of dust content for most galaxies in our
samples. However, it is clear that they do not trace the \emph{same}
dust because the $\tau_V$ measured from the Balmer emission lines
(upper axis) is an order of magnitude lower than that indicated by the
mid-IR silicate absorption strengths (right-hand axis, from zero-age
starburst models of Rowan-Robinson \& Efstathiou 2009).
The dotted line is the best-fit squared relation to the SSGSS-starforming,
composite and \hds\ galaxies:
\begin{equation}
{\rm Sil}_{9.7\um} = -0.17\left(\frac{f(\ha)}{B f(\hb)} -1\right)^2
\end{equation}

In the case of starforming galaxies without a strong AGN, the mid-IR
continuum predominantly arises from dust grains situated within the
\hii\ regions re-radiating the light from young and hot stars. It is
this dust continuum that is absorbed by the Silicate grains along the
line-of-sight between the light source and the observer. Balmer
emission lines also come from ionised gas within \hii\ regions. Thus,
the observed correlation between the Silicate absorption strength and
Balmer decrement implies either: (i) a correlation between the amount
of dust in dense stellar birth clouds which cause the Silicate
absorption, and the amount of dust in more diffuse stellar birth
clouds which cause the absorption of the optical emission lines; or
(ii) a self-similar structure for \hii\ regions in which the Balmer
lines that reach the observer have passed through a fixed fraction of
the dense dust clouds. 

There are two notable exceptions where the optical Balmer decrement
fails to predict the presence of an extremely significant attenuation
in the mid-IR: the ULIRGs 15250 and 08572.  \citet{Nardini:2009p4304}
find that both of these ULIRGs have significant contributions from AGN
to their bolometric luminosity (53 and 87\% respectively), using a
spectral decomposition between 5 and 8\um. Models have shown that
extremely strong silicate absorption, as observed in these two ULIRGs,
requires a steep temperature gradient within the obscuring cloud,
which can only be obtained where the light source responsible for
heating the dust that causes the mid-IR continuum is deeply buried in
a smooth thick shell of dust that is both geometrically and optically
thick \citep{Levenson:2007p2766}. This is again consistent with the
source being an AGN, although a nuclear starburst may suffice. A
clumpy distribution of dust leads to much shallower absorption or
emission \citep{Nenkova:2002p2744}.  Our emission line analysis is
also consistent with these galaxies harbouring a buried AGN, they have
an increased [Ne]/\ha\ ratio and lie in the composite region of the
BPT diagram. While the position of these galaxies on the BPT diagram
is not in itself conclusive proof that these two ULIRGs harbour an
AGN, if their position is caused by the AGN then it indicates that
ionising radiation from the AGN escapes to ionise the NLR, despite the
dense dust cloud that presumably surrounds the central continuum
source. Two dusty Seyfert galaxies and one ULIRG with a Seyfert
nucleus (Mrk273) also lie above the relation defined by the SSGSS and
\hds\ galaxies, further supporting the idea that dust along the
line-of-sight to a deeply buried AGN can lead to stronger silicate
absorption, relative to nebular line absorption, than normal \hii\
regions.

We are not aware of any models which simultaneously predict the
strength of silicate absorption and attenuation of the optical
emission lines. Qualitatively Figure \ref{fig:dust} suggests a two
component model is required to describe the dust distribution in
galaxies/AGN. The lower branch is consistent with silicate absorption
arising in the clumpy distribution of stellar birthclouds, with the
densest clouds causing some silicate absorption, and the Balmer
emission lines originating from the same star formation regions but
further out in the clouds. The upwards scatter of some sources from
this branch is consistent with the dust being concentrated in a dense
shell-like distribution obscuring a single nuclear source, most likely
an AGN in the cases studied here, but also conceivably a strong
nuclear starburst.

\subsection{Galaxies with an AGN}\label{sec:bhar}

\begin{figure*}
\includegraphics[scale=0.5]{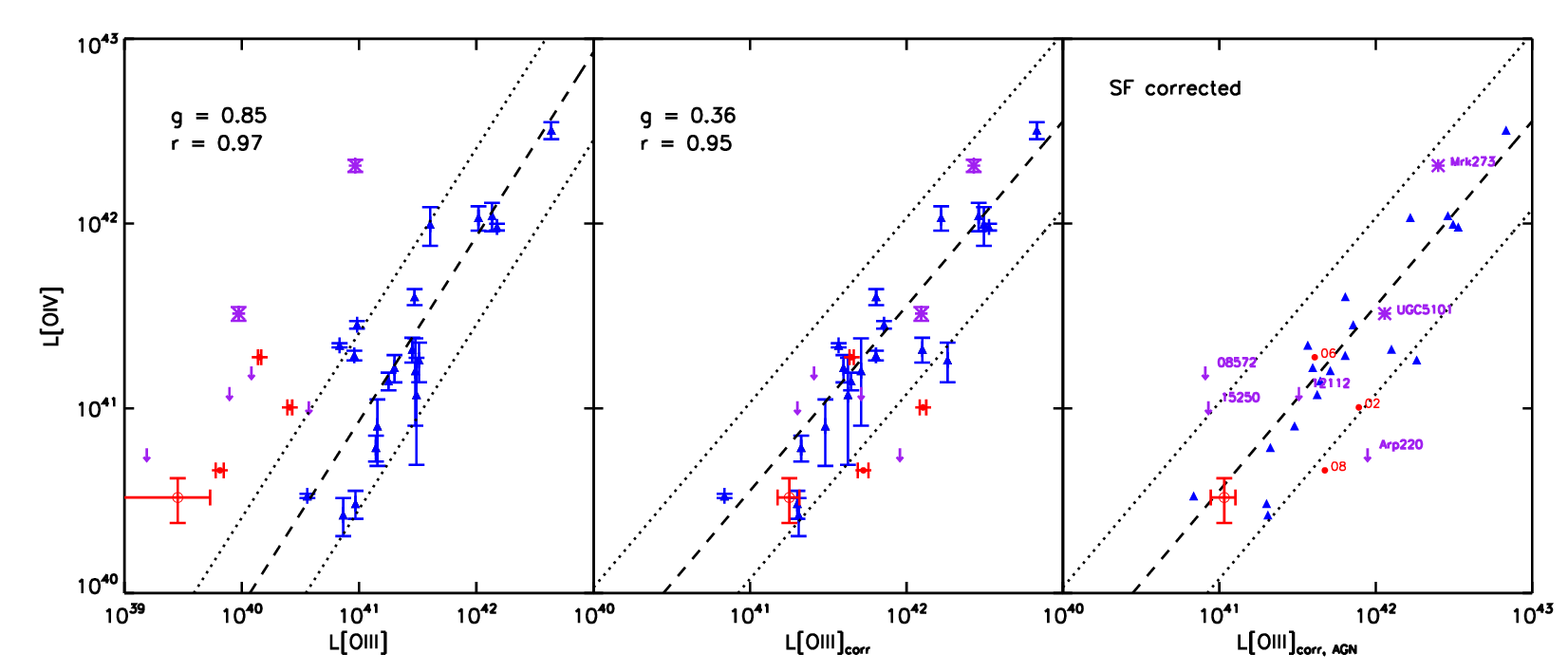}
\caption{\oiiilong\ vs. \oivlong\ luminosities (erg/s) for the Seyfert
  sample (filled blue triangles, 20/20), ULIRGs (purple stars, 2/6),
  and the \hds\ galaxies (red circles, 3/11) with \oiv, \oiii, \ha\ and
  \hb\ detected at $>3\sigma$. The stack of the remaining \hds\
  galaxies with non-detections in \oiv\ is shown as a red open
  circle. {\it Left:} \loiii\ is not corrected for dust attenuation,
  or for contamination due to star formation. Note the different
  x-axis range in this panel. {\it Center:} \loiii\ is corrected for
  dust attenuation but not for contamination due to star
  formation. {\it Right:} \loiii\ of the ULIRGs and \hds\ galaxies has
  been corrected for contamination from star formation using the
  method described in Wild et al. (2010). Dust attenuation is
  calculated using the best-fit dust-curve derived in Section
  \ref{sec:fitdust}. Overplotted as a dashed line is the best-fit
  linear relation to the Seyfert sample only, the dotted lines
  indicate a factor of three above and below this relation. For
  clarity, error bars are omitted in the right hand
  panel. }\label{fig:oiiioiv}
\end{figure*}

As explained in Section \ref{sec:nir}, a dust curve and accurate
Balmer decrement that adequately corrects \hii\ region emission lines
for dust attenuation, will not necessarily do the same for lines from
the NLR of AGN. Particularly in composite objects, a higher fraction
of Balmer lines will arise from within the \hii\ regions of the galaxy
compared to lines such as \oiii\ which are stronger in NLRs.
Therefore a form of differential extinction might arise between the
Balmer and higher ionisation lines, and a corresponding bias in the
correction of AGN NLR lines for dust using the measured Balmer
decrement. Additionally, the form of the dust curve may not be the
same for light originating from \hii\ regions and from an AGN NLR, due
to the different relative geometry between the light source and the
dust cloud, and the different balance between dense and diffuse dust
attenuating the lines. The mid-IR \oiv\ line at 25.89 \mum\ has been
shown to be an accurate indicator of AGN power
\citep{2008ApJ...682...94M}, and the \oiii/\oiv\ ratio a possible
measure of AGN attenuation \citep{2005A&A...442L..39H}. In this
section we test the accuracy of dust-corrected \loiii\ as a measure of
AGN luminosity in our sources by comparing to the mid-IR \oiv\ line.

In Figure \ref{fig:oiiioiv} we plot dust-uncorrected and
dust-corrected \loiii\ vs. \loivlong\ for the Seyferts, the 3 \hds\
galaxies and the 2 ULIRGs with measured \oiv\ lines. Note that the
correlation coefficient is similar for the Seyferts whether the dust
correction is performed or not. These Seyferts have very small Balmer
decrements so the dust correction has only a small effect.  Because
some of these objects are classified as composite AGN-starformation by
their optical line ratios, and \oiii\ is also produced by
starformation, in the right hand panel we apply a correction for this
following the method of \citet{Wild:2010p4207}, based on the expected
\oiii/\ha\ ratio for metal-rich starforming galaxies\footnote{We note
  that such a correction would not be appropriate for the ULIRGs if
  shocks were a significant contributor to the total \ha\
  luminosity.}. \oiv\ is also produced in \hii\ regions, although the
contamination is expected to be much smaller than for \oiii\ due to
the higher ionisation level. Clearly, there is a tight correlation
between dust-corrected \loiii\ and \loiv\ for the Seyfert sample
\citep[see also][]{2010arXiv1007.0900L}. All of the dusty \hds\
galaxies and ULIRGs with measured \oiv\ fall within a factor of 3 of
the relation defined by the Seyferts. Only the upper limit for Arp220
is marginally inconsistent with the relation defined by the Seyfert
galaxies.  Taking the remainder of the \hds\ galaxies and stacking the
SH spectra in luminosity units shows a clear \oiv\ line, which we fit
together with the neighbouring \feii\ line using a double
Gaussian. The corresponding stacked \loiv-\loiii\ luminosity also lies
exactly on the Seyfert relation, once contamination of \loiii\ from
star formation has been accounted for. Errors are estimated from the
standard deviation of the luminosity on removing one spectrum at a
time from the stack. This strong correlation between \oiv\ and \oiii\
dust corrected using the observed Balmer decrement and dust-curve
appropriate for \hii\ regions, for even dusty composite galaxies,
shows that there is no evidence for significant differential
extinction between lines emitted from the NLR and \hii\ regions of a
galaxy. Although slightly surprising, this may imply that the majority
of dust attenuation suffered by NLR emission lines comes from the
diffuse dust and \hii\ regions in the intervening galaxy, through
which they must pass on their way to the observer.

Unfortunately neither of the two ULIRGs with extreme silicate
absorption (08572 and 15250) have measured \loiv, so we cannot test
directly whether optical and mid-IR NLR lines agree even in the
presence of extreme, dense obscuration of the central light
source. Such a study may be possible with the large samples of ULIRGs
now available in archives.

\section{Discussion}\label{sec:disc} 

The potential of dust attenuation and complicated dust geometries to
render useless optical observations of extreme galaxies is sometimes
used to justify rest-frame IR observations. In the previous sections
we have used the correlation between the unattenuated mid-IR and
attenuated optical nebular emission lines to argue that a single dust
curve can accurately correct optical emission lines for dust
attenuation in a large range of different galaxies. The correction is
accurate to within a factor of 2 in most cases, and to at least a
factor of 4 for even the most heavily obscured ULIRGs.

Here, we discuss the implications of our work for deriving SFRs and
black hole accretion rates from the optical emission lines \ha, \oii,
and \oiii. We present useful relations for dust corrections at the
wavelengths of each of these lines.

\subsection{Star formation rates}\label{sec:sfr}

\begin{figure*}
\includegraphics[scale=0.5]{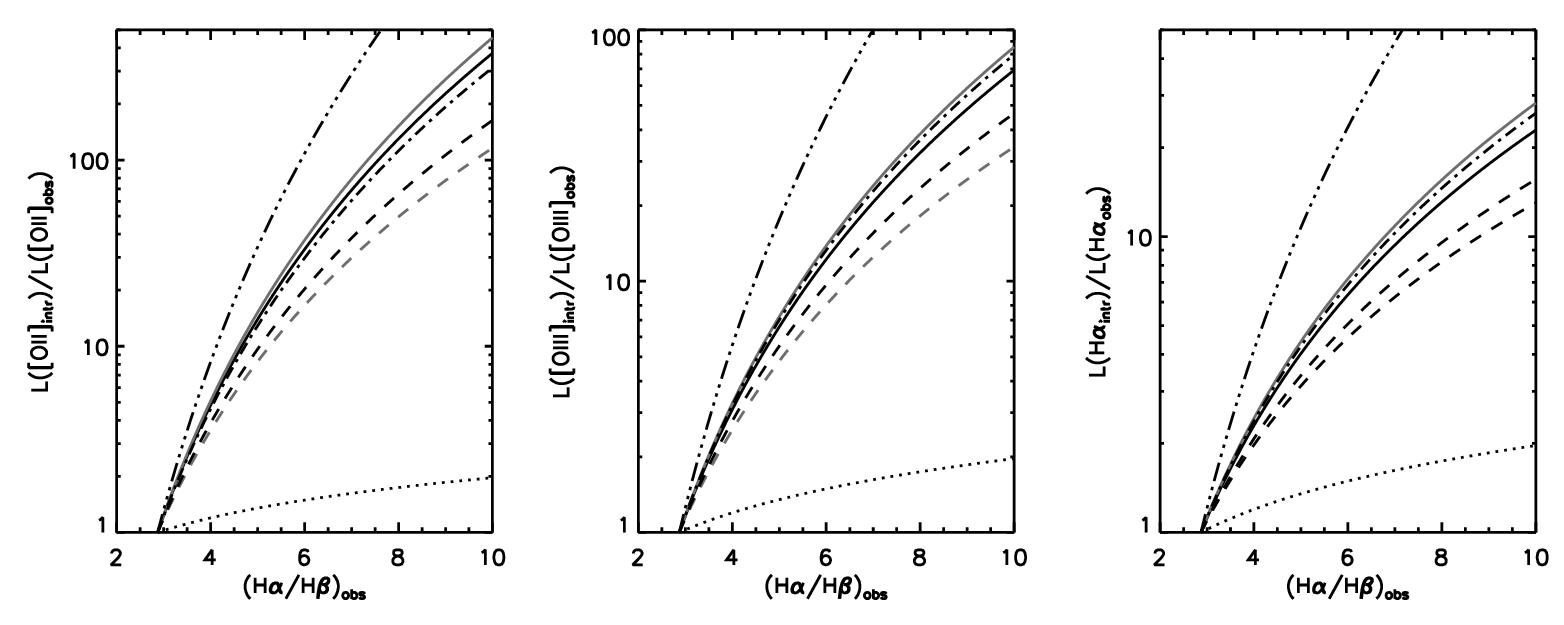}
\caption{Intrinsic to observed flux ratios as a function of observed
  \ha/\hb\ line ratio for different dust curves, at wavelengths
  corresponding to \oiilong (left), \oiiilong\ (center) and \ha\
  (right).  The lines are (as given in Figure \ref{fig:dustlaws}): the
  SMC extinction curve with an R$_V=3.1$
  \citep[dotted][]{Pei:1992p4812}; a MW extinction curve with
  R$_V=3.1$ \citep[dashed black,][]{1994ApJ...422..158O}; a different MW
  extinction curve with R$_V=3.2$ \citep[dashed
  grey,][]{Seaton:1979p4492,Nandy:1975p4502}\; the $\mu_{0.3}$ and
  $\mu_{0.4}$ attenuation curve for emission lines \citep[solid black
  and grey,][]{wild_psb}; the Calzetti attenuation curve for starburst
  continua with R$_V=4.05$ \citep[dash-dot][]{Calzetti:2000p4473}; the
  $\lambda^{-0.7}$ attenuation curve for starburst continua from CF00
  (dash-triple-dot). \emph{Note the changing scale on the
    y-axis.}}\label{fig:atten}
\end{figure*}

The left and right panels of Fig. \ref{fig:atten} show the fractional
attenuation (in flux units) as a function of observed Balmer decrement
at \ha\ and \oii, for each of the different dust curves discussed in
Section \ref{sec:results}.  Using the best-fit $\mu_{0.4}$ dust curve,
the recommended correction at \ha\ is:
\begin{equation}
{\rm L[H\alpha]_{corr}} ={\rm L[H\alpha]_{obs}}\left(\frac{\rm (H\alpha/H\beta)_{obs}}{B}\right)^{2.67_{-0.32}^{+0.40}} 
\end{equation}
and at \oii\footnote{Note that this is an extrapolation of the
  dust-curve beyond the wavelength region directly tested in this
  paper.}:
\begin{equation}
{\rm L[OII]_{corr}} ={\rm L[OII]_{obs}}\left(\frac{\rm (H\alpha/H\beta)_{obs}}{B}\right)^{4.89_{-0.29}^{+0.36}} 
\end{equation}
where $B$ is the unattenuated ratio as discussed above and errors are
propagated from the errors on the best-fit $\mu$. While for \hii\
galaxies and strong Seyferts the choice of $B$ is clear, for composite
galaxies the unknown $B$ may cause small systematic
uncertainties. However as \hii\ regions have relatively stronger \ha\
emission, a value of $B=2.86$ is probably close to reality.

To derive a star formation rate from optical nebular emission lines,
two steps are required. This paper has focused on the first step: to
correct the line for dust attenuation. The second step is to convert
the dust-corrected line luminosity into a star formation rate.  While
our results show that dust-corrected \ha\ luminosity traces [Ne]
luminosity, this is not enough to confirm the reliability of \ha\ as
an indicator of star formation rate in extremely dusty objects. This
is because dense dust clouds can have high optical depths even in the
mid-IR, as indicated by the strong Silicate absorption in some
galaxies, and thus even the neon lines may be attenuated in some
objects. In addition, the most extreme compact, dusty galaxies may
even have a non-negligible fraction of their ionizing photons absorbed
by dust and not gas (and thus not traced by H$\alpha$). Encouragingly,
Ho \& Keto (2007) have already shown how [Ne] luminosity correlates
strongly with both the total infrared luminosity and Brackett-$\alpha$
luminosity in star-forming galaxies. Their galaxy sample includes a
handful of ULIRGs, showing that these relations hold even at extreme
luminosities. This result provides circumstantial evidence that the
neon luminosity is tracing star-formation even in the dustiest of
galaxies.  In contrast, Farrah et al. (2007) find [Ne] luminosity in
ULIRGs to be deficient by 0.4dex from the Ho \& Keto relation, which
they ascribe to extinction of the [Ne] lines. In a companion paper we
plan to address this issue directly by comparing the star formation
rates derived from the full SEDs, with that derived from the emission
lines.

\subsection{Black hole accretion rates}\label{sec:bolcorrn}

As with star formation rate, the derivation of black hole accretion
rates (BHAR) from nebular emission line luminosities requires two
steps: correction for dust emission discussed in this paper, and then
conversion into BHAR via a bolometric correction.

For the best-fit $\mu_{0.4}$ dust curve, the relevant correction for
dust attenuation at \oiii\ is:
\begin{equation}
{\rm L[OIII]_{corr}} ={\rm L[OIII]}\left(\frac{\rm (H\alpha/H\beta)_{obs}}{B}\right)^{3.55_{-0.32}^{+0.40}} 
\end{equation} 
where B is the unattenuated ratio as discussed above.  The correction
caused by the different dust curves is presented in the central panel
of Figure \ref{fig:atten}.

The exponent for the $\mu_{0.4}$ dust curve is larger than that used
in some AGN literature and can lead to significant differences in
estimated bolometric AGN luminosity and accretion rates as a fraction
of $L_{\rm Edd}$ \citep[see also the discussion
in][]{Netzer:2009p4406}.  Using a smaller exponent than appropriate
will cause an underestimate in \loiii$_{\rm corr}$ which will vary
with dust content. For example, for the relatively dust-free sample of
Seyfert galaxies studied in this work, the difference between the
$\mu_{0.4}$ dust curve exponent of 3.55, and 2.94 (from a MW
extinction law) used in e.g. \citet{Bassani:1999p4370} and
\citet{Lamastra:2009p4236} leads to differences ranging from 5 to
30\%. The potential increase in the under- (or over-)estimation of the
line luminosity with increasing dust content because of an
inappropriate dust-curve has important implications for understanding
whether the bolometric correction for \loiii\ changes systematically
with \loiii, due to the correlations between SFR of the host galaxy
and the luminosity of the central AGN
\citep[e.g.][]{2003MNRAS.346.1055K, Schweitzer:2006p4405,
  Zheng:2009p4373, Netzer:2009p4406} and the SFR and dust content of
the host \citep{Cunha:2010p3914}. Our results do not explicitly
constrain which dust-curve is most appropriate for Seyferts, however,
we would like to emphasise the importance of including the errors
caused by the dust-attenuation correction when studying emission line
trends.

\subsection{AGN detection}

Even in the two ULIRGs in our sample where there is evidence for a
heavily buried central nucleus, we would like to emphasise that the
optical emission line ratios still indicate the possible presence of
an AGN, albeit not conclusively. This shows that even where the AGN is
deeply buried, some radiation may still escape to ionise the NLR
clouds. This has important implications for using optical emission
line ratios to identify AGN in objects such as ULIRGs, and will become
increasingly important at high-z. Unfortunately neither of the two
ULIRGs with strong silicate absorption in our sample have measured
\oiv, and therefore we cannot directly verify that their position on
the BPT diagram, and therefore their correction of \oiii\ for
contamination from star formation, is correct. The specific study of a
larger sample of ULIRGs to compare the bolometric AGN luminosities
derived from mid-IR lines, mid-IR continuum and dust- and
star-formation corrected high-quality optical emission lines would now
be possible with samples in the SDSS and Spitzer archives.

\section{Conclusions}\label{sec:conclude}

The purpose of this paper was to verify, or otherwise, the accuracy of
fundamental quantities derived from optical emission lines,
particularly in the case of significant dust attenuation of the
lines. In particular, optical emission lines provide powerful measures
of star formation rate and black hole accretion rate.  We compared two
``dusty'' galaxy samples, including 6 ULIRGs, to a sample of
``ordinary'' galaxies from the SSGSS all with high quality SDSS
optical spectra and mid-IR spectra from the \emph{Spitzer Space
  Telescope} IRS spectrograph. The results are extremely encouraging
for using rest-frame optical spectra to study dusty galaxies with an
$A_V<4.5$. Our results are promising even for galaxies as dusty as
ULIRGs, although a much larger sample should be studied covering the
full range of this diverse population, in order to verify our
conclusions. Our results have important implications for future
high-redshift spectroscopic surveys, where galaxies have higher star
formation rates and dust contents may be correspondingly higher.

We have compared several different dust curves used in the literature
to correct optical emission lines for dust attenuation. Our results
favour a dust curve which lies close to the MW extinction curve,
indicating that a significant fraction of the dust which attenuates
nebular emission lines has a screen-like geometry. This is in
agreement with the model for emission line attenuation presented by
\citet{wild_psb} and \citet{2008MNRAS.388.1595D} and based upon the
2-component CF00 dust model. In this model, the screen-like extinction
arises from the dense stellar birth clouds which surround the hottest
stars responsible for ionising the gas. We find a best-fit dust-curve
with on average 40\% of the optical depth at 5500\AA\ arising from
diffuse ISM dust. The $\mu_{0.4}$ dust curve is not significantly
different from the Calzetti dust-curve in the optical wavelength
regime, however, we focus our results and analysis on the former which
was intended for emission lines and allows greater flexibility, rather
than the latter which was derived from galaxy continua.

Particular conclusions with regard to measuring star formation from
optical emission lines in the presence of dust: 
\begin{itemize}
\item The MW, Calzetti, $\mu_{0.3}$ and best-fit $\mu_{0.4}$ dust
  curves all allow accurate correction of \ha\ (to better than a
  factor of 2), and therefore accurate calculation of star formation
  rates, for galaxies with observed \ha/\hb\ ratios of as much as 10
  ($\tau_V<4.5$).
\item For only 2/6 ULIRGs with $\tau_V<6$ does the optical \ha\ differ
  from the mid-IR [Ne] lines by more than a factor of 2, but then only
  by a factor of 4. In these two cases (08572 and 15250) we have
  argued that the observed excess of [Ne] is in fact caused by a
  buried AGN, rather than uncertainties in the dust correction
  \citep[in agreement with the results of][]{Nardini:2009p4304}. 
\item The majority of galaxies in our samples favour a dust curve
  close to the MW screen-like dust extinction curve. This suggests
  that much of the extra dust in the unusually dusty objects is
  located in dense clouds, rather than in the diffuse ISM.
\item There is tentative evidence for a range in the shape of the dust
  curves in ordinary starforming galaxies. For the CF00 dust model,
  this range translates into a variation in the fraction of optical
  depth arising in the diffuse ISM of between around 30\% and 90\%, with the
  remaining fraction arising in dense birth clouds. However, the
  variation in the intrinsic [Ne]/\ha\ ratio with metallicity, the
  issue of relative aperture correction between the IRS and SDSS data,
  and the large statistical errors on the \hb\ emission line, prevent
  us from quantifying this scatter more precisely.
\end{itemize}

Particular conclusions with regard to identifying AGN and measuring
their accretion rates from optical emission lines in the presence of
dust:
\begin{itemize}
\item The best-fit $\mu_{0.4}$ dust curve and measured Balmer
  decrement allows accurate correction of \oiii\ for dust attenuation,
  to within a factor of 3. This is slightly surprising, and suggests
  that the effects of variations in the relative geometry of the
  source and dust, and differential extinction between the Balmer and
  higher ionisation lines, are minimal.
\item The method to separate the contributions to optical emission
  lines from both star formation and AGN presented in
  \citet{Wild:2010p4205} tightens the relation between \loiv\ and
  \loiii\ for composite objects. This supports the use of the position
  of galaxies on the optical BPT line ratio diagram to measure the
  amount of \loiii\ originating from the AGN, and therefore the
  estimates of black hole accretion rates from \loiii\ corrected
  for dust attenuation and star formation contamination. 
\item All ULIRGs in our sample are classified as AGN or composite
  objects from their optical emission line ratios. While the composite
  line ratios may be caused by shocks, we have argued that other
  observational results favour the presence of a buried AGN
  contributing significantly to the mid-IR continuum. It is therefore
  possible that sufficient ionising flux escapes from the central
  source to ionise the NLR, even in the case of heavily buried nuclei
  such as in ULIRGS 08572 and 15250. Unfortunately these two ULIRGs do
  not have measured \loiv\ preventing us from verifying our
  corrections to \loiii\ for dust attenuation and star formation
  contamination in such extreme cases. In the case of ULIRGs, whose
  properties vary widely, clearly a much larger sample should be
  studied before firm conclusions can be drawn.
\end{itemize}

\section{Acknowledgements}
We would like to thank Jarle Brinchmann for his help in understanding
aperture bias issues and Stephane Charlot for comments on an early
draft. We thank the anonymous referee for a careful reading of the
manuscript and comments which improved its clarity. The function
fitting performed in this paper used the IDL MPFIT software
http://purl.com/net/mpfit (Markwardt 2009). The ITERA package was used
to investigate intrinsic emission line ratios
http://www.strw.leidenuniv.nl/~brent/itera.html
\citep{Groves:2010p4895}.

Funding for the SDSS and SDSS-II has been provided by the Alfred
P. Sloan Foundation, the Participating Institutions, the National
Science Foundation, the U.S. Department of Energy, the National
Aeronautics and Space Administration, the Japanese Monbukagakusho, the
Max Planck Society, and the Higher Education Funding Council for
England. The SDSS Web Site is http://www.sdss.org/. 

The SDSS is managed by the Astrophysical Research Consortium for the
Participating Institutions. The Participating Institutions are the
American Museum of Natural History, Astrophysical Institute Potsdam,
University of Basel, University of Cambridge, Case Western Reserve
University, University of Chicago, Drexel University, Fermilab, the
Institute for Advanced Study, the Japan Participation Group, Johns
Hopkins University, the Joint Institute for Nuclear Astrophysics, the
Kavli Institute for Particle Astrophysics and Cosmology, the Korean
Scientist Group, the Chinese Academy of Sciences (LAMOST), Los Alamos
National Laboratory, the Max-Planck-Institute for Astronomy (MPIA),
the Max-Planck-Institute for Astrophysics (MPA), New Mexico State
University, Ohio State University, University of Pittsburgh,
University of Portsmouth, Princeton University, the United States
Naval Observatory, and the University of Washington.  

\bibliographystyle{mn2e}

\begin{thebibliography}{}

\bibitem[\protect\citeauthoryear{Adelman-McCarthy, et al \&
  collaboration}{Adelman-McCarthy et~al.}{2008}]{AdelmanMcCarthy:2008p4857}
Adelman-McCarthy J.~K.,  et al   collaboration T.~S.,  2008, \apjs, 175, 297

\bibitem[\protect\citeauthoryear{Armus, Charmandaris, Bernard-Salas \& et
  al.}{Armus et~al.}{2007}]{Armus:2007p964}
Armus L.,  Charmandaris V.,  Bernard-Salas J.,    et al. 2007, \apj, 656, 148

\bibitem[\protect\citeauthoryear{Baldwin, Phillips \& Terlevich}{Baldwin
  et~al.}{1981}]{1981PASP...93....5B}
Baldwin J.~A.,  Phillips M.~M.,    Terlevich R.,  1981, \pasp, 93, 5

\bibitem[\protect\citeauthoryear{Bassani, Dadina, Maiolino, Salvati, Risaliti,
  della Ceca, Matt \& Zamorani}{Bassani et~al.}{1999}]{Bassani:1999p4370}
Bassani L.,  Dadina M.,  Maiolino R.,  Salvati M.,  Risaliti G.,  della Ceca
  R.,  Matt G.,    Zamorani G.,  1999, \apjs, 121, 473

\bibitem[\protect\citeauthoryear{B{\"o}ker, Calzetti, Sparks, Axon, Bergeron,
  Bushouse, Colina, Daou, Gilmore, Holfeltz, MacKenty, Mazzuca, Monroe, Najita,
  Noll, Nota, Ritchie, Schultz, Sosey, Storrs \& Suchkov}{B{\"o}ker
  et~al.}{1999}]{Boker:1999p4704}
B{\"o}ker T.,  Calzetti D.,  Sparks W.,  Axon D.,  Bergeron L.~E.,  Bushouse
  H.,  Colina L.,  Daou D.,  Gilmore D.,  Holfeltz S.,  MacKenty J.,  Mazzuca
  L.,  Monroe B.,  Najita J.,  Noll K.,  Nota A.,  Ritchie C.,  Schultz A.,
  Sosey M.,  Storrs A.,    Suchkov A.,  1999, \apjs, 124, 95

\bibitem[\protect\citeauthoryear{Borgne, Elbaz, Ocvirk \& Pichon}{Borgne
  et~al.}{2009}]{LeBorgne:2009p4512}
Borgne D.~L.,  Elbaz D.,  Ocvirk P.,    Pichon C.,  2009, \aap, 504, 727

\bibitem[\protect\citeauthoryear{Brinchmann, Charlot, White, Tremonti,
  Kauffmann, Heckman \& Brinkmann}{Brinchmann
  et~al.}{2004}]{2004MNRAS.351.1151B}
Brinchmann J.,  Charlot S.,  White S.~D.~M.,  Tremonti C.,  Kauffmann G.,
  Heckman T.,    Brinkmann J.,  2004, \mnras, 351, 1151

\bibitem[\protect\citeauthoryear{Bruzual \& Charlot}{Bruzual \&
  Charlot}{2003}]{2003MNRAS.344.1000B}
Bruzual G.,  Charlot S.,  2003, \mnras, 344, 1000

\bibitem[\protect\citeauthoryear{Calzetti}{Calzetti}{2001}]{2001PASP..113.1449%
C}
Calzetti D.,  2001, \pasp, 113, 1449

\bibitem[\protect\citeauthoryear{Calzetti, Armus, Bohlin, Kinney, Koornneef \&
  Storchi-Bergmann}{Calzetti et~al.}{2000}]{Calzetti:2000p4473}
Calzetti D.,  Armus L.,  Bohlin R.~C.,  Kinney A.~L.,  Koornneef J.,
  Storchi-Bergmann T.,  2000, \apj, 533, 682

\bibitem[\protect\citeauthoryear{Calzetti, Kinney \& Storchi-Bergmann}{Calzetti
  et~al.}{1994}]{1994ApJ...429..582C}
Calzetti D.,  Kinney A.~L.,    Storchi-Bergmann T.,  1994, \apj, 429, 582

\bibitem[\protect\citeauthoryear{Calzetti, Kinney \& Storchi-Bergmann}{Calzetti
  et~al.}{1996}]{Calzetti:1996p4710}
Calzetti D.,  Kinney A.~L.,    Storchi-Bergmann T.,  1996, \apj. 458, 132

\bibitem[\protect\citeauthoryear{Cardelli, Clayton \& Mathis}{Cardelli
  et~al.}{1989}]{1989ApJ...345..245C}
Cardelli J.~A.,  Clayton G.~C.,    Mathis J.~S.,  1989, \apj, 345, 245

\bibitem[\protect\citeauthoryear{Charlot \& Fall}{Charlot \&
  Fall}{2000}]{2000ApJ...539..718C}
Charlot S.,  Fall S.~M.,  2000, \apj, 539, 718 (CF00)

\bibitem[\protect\citeauthoryear{da Cunha, Charlot \& Elbaz}{da~Cunha
  et~al.}{2008}]{2008MNRAS.388.1595D}
da Cunha E.,  Charlot S.,    Elbaz D.,  2008, \mnras, 388, 1595

\bibitem[\protect\citeauthoryear{da Cunha, Eminian, Charlot \&
  Blaizot}{da~Cunha et~al.}{2010}]{Cunha:2010p3914}
da Cunha E.,  Eminian C.,  Charlot S.,    Blaizot J.,  2010, arXiv, astro-ph.CO


\bibitem[\protect\citeauthoryear{D{\'\i}az-Santos, Alonso-Herrero, Colina,
  Packham, Levenson, Pereira-Santaella, Roche \& Telesco}{D{\'\i}az-Santos
  et~al.}{2010}]{DiazSantos:2010p5114}
D{\'\i}az-Santos T.,  Alonso-Herrero A.,  Colina L.,  Packham C.,  Levenson
  N.~A.,  Pereira-Santaella M.,  Roche P.~F.,    Telesco C.~M.,  2010,
  \apj, 711, 328

\bibitem[\protect\citeauthoryear{Eliche-Moral, Prieto, Gallego \&
  Zamorano}{Eliche-Moral et~al.}{2010}]{ElicheMoral:2010p5018}
Eliche-Moral M.~C.,  Prieto M.,  Gallego J.,    Zamorano J.,  2010, eprint
  arXiv, 1003, 686

\bibitem[\protect\citeauthoryear{Farrah et al.}{2007}]{2007ApJ...667..149F} 
Farrah D., et al., 2007, ApJ, 667, 149 

\bibitem[\protect\citeauthoryear{Fernandes, Stasi{\'n}ska, Schlickmann, Mateus,
  Asari, Schoenell \& Sodr{\'e}}{Fernandes
  et~al.}{2010}]{CidFernandes:2010p4744}
Fernandes R.~C.,  Stasi{\'n}ska G.,  Schlickmann M.~S.,  Mateus A.,  Asari
  N.~V.,  Schoenell W.,    Sodr{\'e} L.,  2010, \mnras, 403, 1036

\bibitem[\protect\citeauthoryear{Groves \& Allen}{Groves \&
  Allen}{2010}]{Groves:2010p4895}
Groves B.~A., Allen M.~G., 2010, New Astronomy, 15, 614 

\bibitem[\protect\citeauthoryear{Groves, Dopita \& Sutherland}{Groves
  et~al.}{2004}]{Groves:2004p4703}
Groves B.~A.,  Dopita M.~A.,    Sutherland R.~S.,  2004, \apjs, 153, 75

\bibitem[\protect\citeauthoryear{Groves, Heckman \& Kauffmann}{Groves
  et~al.}{2006}]{Groves:2006p4862}
Groves B.~A.,  Heckman T.~M.,    Kauffmann G.,  2006, \mnras, 371, 1559

\bibitem[Haas et 
al.(2005)]{2005A&A...442L..39H} Haas, M., Siebenmorgen, R., Schulz,
B., Kr{\"u}gel, E., \& Chini, R.\ 2005, \aap, 442, L39  


\bibitem[\protect\citeauthoryear{Heckman, Kauffmann, Brinchmann, Charlot,
  Tremonti \& White}{Heckman et~al.}{2004}]{heckman04}
Heckman T.~M.,  Kauffmann G.,  Brinchmann J.,  Charlot S.,  Tremonti C.,
  White S.~D.~M.,  2004, \apj, 613, 109

\bibitem[\protect\citeauthoryear{Higdon, Armus, Higdon, Soifer \& Spoon}{Higdon
  et~al.}{2006}]{Higdon:2006p4736}
Higdon S. J.~U.,  Armus L.,  Higdon J.~L.,  Soifer B.~T.,    Spoon H. W.~W.,
  2006, \apj, 648, 323

\bibitem[\protect\citeauthoryear{Higdon, Devost, Higdon, Brandl, Houck, Hall,
  Barry, Charmandaris, Smith, Sloan \& Green}{Higdon
  et~al.}{2004}]{Higdon:2004p2644}
Higdon S. J.~U.,  Devost D.,  Higdon J.~L.,  Brandl B.~R.,  Houck J.~R.,  Hall
  P.,  Barry D.,  Charmandaris V.,  Smith J. D.~T.,  Sloan G.~C.,    Green J.,
  2004, \pasp, 116, 975

\bibitem[\protect\citeauthoryear{Ho \& Keto}{Ho \& Keto}{2007}]{Ho:2007p4951}
Ho L.~C.,  Keto E.,  2007, \apj, 658, 314

\bibitem[\protect\citeauthoryear{Hopkins, Hernquist, Cox, Matteo, Robertson \&
  Springel}{Hopkins et~al.}{2006}]{2006ApJS..163....1H}
Hopkins P.~F.,  Hernquist L.,  Cox T.~J.,  Matteo T.~D.,  Robertson B.,
  Springel V.,  2006, \apjs, 163, 1

\bibitem[\protect\citeauthoryear{Houck, Roellig, van Cleve \& et al.}{Houck
  et~al.}{2004}]{Houck:2004p2735}
Houck J.~R.,  Roellig T.~L.,  van Cleve J.,    et al. 2004, \apjs, 154, 18

\bibitem[\protect\citeauthoryear{Ilbert, Salvato, Floc'h \& et al.}{Ilbert
  et~al.}{2010}]{Ilbert:2010p5037}
Ilbert O.,  Salvato M.,  Floc'h E.~L.,    et al. 2010, \apj, 709, 644

\bibitem[\protect\citeauthoryear{Johnson, Schiminovich, O'Dowd, Meder \&
  Treyer}{Johnson et~al.}{2009}]{Johnson:2009p2742}
Johnson B.~D.,  Schiminovich D.,  O'Dowd M.,  Meder K.,    Treyer M.,  2009,
  The Evolving ISM in the Milky Way and Nearby Galaxies, p.~31

\bibitem[\protect\citeauthoryear{Kauffmann, Heckman, Tremonti \& et
  al}{Kauffmann et~al.}{2003}]{2003MNRAS.346.1055K}
Kauffmann G.,  Heckman T.~M.,  Tremonti C.,    et al 2003, \mnras, 346, 1055

\bibitem[\protect\citeauthoryear{Kauffmann, Heckman, White \& et al}{Kauffmann
  et~al.}{2003}]{2003MNRAS.341...33K}
Kauffmann G.,  Heckman T.~M.,  White S.~D.~M.,    et al 2003, \mnras, 341, 33

\bibitem[\protect\citeauthoryear{Kennicutt, Hao, Calzetti, Moustakas, Dale,
  Bendo, Engelbracht, Johnson \& Lee}{Kennicutt
  et~al.}{2009}]{Kennicutt:2009p4470}
Kennicutt R.~C.,  Hao C.-N.,  Calzetti D.,  Moustakas J.,  Dale D.~A.,  Bendo
  G.,  Engelbracht C.~W.,  Johnson B.~D.,    Lee J.~C.,  2009, \apj, 703, 1672

\bibitem[\protect\citeauthoryear{Kewley, Dopita, Sutherland, Heisler \&
  Trevena}{Kewley et~al.}{2001}]{2001ApJ...556..121K}
Kewley L.~J.,  Dopita M.~A.,  Sutherland R.~S.,  Heisler C.~A.,    Trevena J.,
  2001, \apj, 556, 121

\bibitem[Kewley et al.(2006)]{2006MNRAS.372..961K} Kewley, L.~J., Groves, 
B., Kauffmann, G., \& Heckman, T.\ 2006, \mnras, 372, 961 

\bibitem[\protect\citeauthoryear{LaMassa, Heckman, Ptak, Hornschemeier,
  Martins, Sonnentrucker \& Tremonti}{LaMassa et~al.}{2009}]{LaMassa:2009p5102}
LaMassa S.~M.,  Heckman T.~M.,  Ptak A.,  Hornschemeier A.,  Martins L.,
  Sonnentrucker P.,    Tremonti C.,  2009, \apj, 705, 568

\bibitem[LaMassa et al.(2010)]{2010arXiv1007.0900L} LaMassa, S.~M., 
Heckman, T.~M., Ptak, A., Martins, L., Wild, V., Sonnentrucker, P., 
\& Tremonti, C.\ 2010, arXiv:1007.0900 

\bibitem[\protect\citeauthoryear{Lamastra, Bianchi, Matt, Perola, Barcons \&
  Carrera}{Lamastra et~al.}{2009}]{Lamastra:2009p4236}
Lamastra A.,  Bianchi S.,  Matt G.,  Perola G.~C.,  Barcons X.,    Carrera
  F.~J.,  2009, \aap, 504, 73

\bibitem[\protect\citeauthoryear{Levenson, Sirocky, Hao, Spoon, Marshall,
  Elitzur \& Houck}{Levenson et~al.}{2007}]{Levenson:2007p2766}
Levenson N.~A.,  Sirocky M.~M.,  Hao L.,  Spoon H. W.~W.,  Marshall J.~A.,
  Elitzur M.,    Houck J.~R.,  2007, \apj, 654, L45

\bibitem[\protect\citeauthoryear{Levesque, Kewley \& Larson}{Levesque
  et~al.}{2010}]{Levesque:2010p4655}
Levesque E.~M.,  Kewley L.~J.,    Larson K.~L.,  2010, \aj, 139, 712

\bibitem[Markwardt(2009)]{2009ASPC..411..251M} Markwardt, C.~B.\ 2009, 
Astronomical Society of the Pacific Conference Series, 411, 251

\bibitem[Mel{\'e}ndez et al.(2008)]{2008ApJ...682...94M} Mel{\'e}ndez, M., 
et al.\ 2008, \apj, 682, 94 

\bibitem[\protect\citeauthoryear{Muzzin, Marchesini, van Dokkum, Labb{\'e},
  Kriek \& Franx}{Muzzin et~al.}{2009}]{Muzzin:2009p4926}
Muzzin A.,  Marchesini D.,  van Dokkum P.~G.,  Labb{\'e} I.,  Kriek M.,
  Franx M.,  2009, \apj, 701, 1839

\bibitem[\protect\citeauthoryear{Naab, Khochfar \& Burkert}{Naab
  et~al.}{2006}]{2006ApJ...636L..81N}
Naab T.,  Khochfar S.,    Burkert A.,  2006, \apjl, 636, L81

\bibitem[\protect\citeauthoryear{Nandy, Thompson, Jamar, Monfils \&
  Wilson}{Nandy et~al.}{1975}]{Nandy:1975p4502}
Nandy K.,  Thompson G.~I.,  Jamar C.,  Monfils A.,    Wilson R.,  1975,
 \aap, 44, 195

\bibitem[\protect\citeauthoryear{Nardini, Risaliti, Salvati, Sani, Watabe,
  Marconi \& Maiolino}{Nardini et~al.}{2009}]{Nardini:2009p4304}
Nardini E.,  Risaliti G.,  Salvati M.,  Sani E.,  Watabe Y.,  Marconi A.,
  Maiolino R.,  2009, \mnras, 399,
  1373

\bibitem[\protect\citeauthoryear{Narron, Ogle \& Laher}{Narron
  et~al.}{2007}]{Narron:2007p4861}
Narron R.,  Ogle P.,    Laher R.~R.,  2007, Astronomical Data Analysis Software
  and Systems XVI ASP Conference Series, 376, 437

\bibitem[\protect\citeauthoryear{Nenkova, Ivezi{\'c} \& Elitzur}{Nenkova
  et~al.}{2002}]{Nenkova:2002p2744}
Nenkova M.,  Ivezi{\'c} {\v Z}.,    Elitzur M.,  2002, \apj, 570, L9

\bibitem[\protect\citeauthoryear{Netzer}{Netzer}{2009}]{Netzer:2009p4406}
Netzer H.,  2009, \mnras, 399, 1907

\bibitem[\protect\citeauthoryear{O'Donnell}{O'Donnell}{1994}]{1994ApJ...422..1%
58O}
O'Donnell J.~E.,  1994, \apj, 422, 158

\bibitem[\protect\citeauthoryear{O'Dowd, Schiminovich, Johnson, Treyer, Martin,
  Wyder, Charlot, Heckman, Martins, Seibert \& van~der Hulst}{O'Dowd
  et~al.}{2009}]{ODowd:2009p4395}
O'Dowd M.~J.,  Schiminovich D.,  Johnson B.~D.,  Treyer M.~A.,  Martin C.~D.,
  Wyder T.~K.,  Charlot S.,  Heckman T.~M.,  Martins L.~P.,  Seibert M.,
  van~der Hulst J.~M.,  2009, \apj, 705, 885

\bibitem[\protect\citeauthoryear{Osterbrock \& Ferland}{Osterbrock \&
  Ferland}{2006}]{Osterbrock:2006p4475}
Osterbrock D.~E.,  Ferland G.~J.,  2006, Astrophysics of gaseous nebulae and
  active galactic nuclei

\bibitem[\protect\citeauthoryear{Pei}{Pei}{1992}]{Pei:1992p4812}
Pei Y.~C.,  1992, \apj, 395, 130

\bibitem[\protect\citeauthoryear{Poggianti \& Wu}{Poggianti \&
  Wu}{2000}]{2000ApJ...529..157P}
Poggianti B.~M.,  Wu H.,  2000, \apj, 529, 157

\bibitem[\protect\citeauthoryear{Rocha, Jonsson, Primack \& Cox}{Rocha
  et~al.}{2008}]{Rocha:2008p4515}
Rocha M.,  Jonsson P.,  Primack J.~R.,    Cox T.~J.,  2008, \mnras, 383, 1281

\bibitem[\protect\citeauthoryear{Rowan-Robinson 
\& Efstathiou}{2009}]{2009MNRAS.399..615R} Rowan-Robinson M., Efstathiou A., 2009, \mnras, 399, 615 

\bibitem[\protect\citeauthoryear{Salim, Charlot, Rich, Kauffmann, Heckman,
  Barlow, Bianchi, Byun \& et al}{Salim et~al.}{2005}]{2005ApJ...619L..39S}
Salim S.,  Charlot S.,  Rich R.~M.,  Kauffmann G.,  Heckman T.~M.,  Barlow
  T.~A.,  Bianchi L.,  Byun Y.-I.,    et al 2005, \apjl, 619, L39

\bibitem[\protect\citeauthoryear{S{\'a}nchez-Bl{\'a}zquez, Peletier,
  Jim{\'e}nez-Vicente, Cardiel, Cenarro, Falc{\'o}n-Barroso, Gorgas, Selam \&
  Vazdekis}{S{\'a}nchez-Bl{\'a}zquez et~al.}{2006}]{miles}
S{\'a}nchez-Bl{\'a}zquez P.,  Peletier R.~F.,  Jim{\'e}nez-Vicente J.,  Cardiel
  N.,  Cenarro A.~J.,  Falc{\'o}n-Barroso J.,  Gorgas J.,  Selam S.,
  Vazdekis A.,  2006, \mnras, 371, 703

\bibitem[\protect\citeauthoryear{Sanders, Soifer, Elias, Madore, Matthews,
  Neugebauer \& Scoville}{Sanders et~al.}{1988}]{Sanders:1988p3844}
Sanders D.~B.,  Soifer B.~T.,  Elias J.~H.,  Madore B.~F.,  Matthews K.,
  Neugebauer G.,    Scoville N.~Z.,  1988, \apj 325, 74

\bibitem[\protect\citeauthoryear{Schweitzer, Lutz, Sturm, Contursi, Tacconi,
  Lehnert, Dasyra, Genzel, Veilleux, Rupke, Kim, Baker, Netzer, Sternberg,
  Mazzarella \& Lord}{Schweitzer et~al.}{2006}]{Schweitzer:2006p4405}
Schweitzer M.,  Lutz D.,  Sturm E.,  Contursi A.,  Tacconi L.~J.,  Lehnert
  M.~D.,  Dasyra K.~M.,  Genzel R.,  Veilleux S.,  Rupke D.,  Kim D.-C.,  Baker
  A.~J.,  Netzer H.,  Sternberg A.,  Mazzarella J.,    Lord S.,  2006, \apj, 649, 79

\bibitem[\protect\citeauthoryear{Seaton}{Seaton}{1979}]{Seaton:1979p4492}
Seaton M.~J.,  1979, \mnras, 187, 73P

\bibitem[\protect\citeauthoryear{Silva, Granato, Bressan \& Danese}{Silva
  et~al.}{1998}]{Silva:1998p4065}
Silva L.,  Granato G.~L.,  Bressan A.,    Danese L.,  1998, \apj, 509, 103

\bibitem[\protect\citeauthoryear{Sirocky, Levenson, Elitzur, Spoon \&
  Armus}{Sirocky et~al.}{2008}]{Sirocky:2008p4863}
Sirocky M.~M.,  Levenson N.~A.,  Elitzur M.,  Spoon H. W.~W.,    Armus L.,
  2008, \apj, 678, 729

\bibitem[\protect\citeauthoryear{Smith, Draine, Dale \& et al.}{Smith
  et~al.}{2007}]{Smith:2007p949}
Smith J. D.~T.,  Draine B.~T.,  Dale D.~A.,    et al. 2007, \apj, 656, 770

\bibitem[\protect\citeauthoryear{Soifer, Sanders, Madore, Neugebauer,
  Danielson, Elias, Lonsdale \& Rice}{Soifer et~al.}{1987}]{Soifer:1987p2581}
Soifer B.~T.,  Sanders D.~B.,  Madore B.~F.,  Neugebauer G.,  Danielson G.~E.,
  Elias J.~H.,  Lonsdale C.~J.,    Rice W.~L.,  1987, \apj, 320, 238

\bibitem[\protect\citeauthoryear{Spoon, Marshall, Houck, Elitzur, Hao, Armus,
  Brandl \& Charmandaris}{Spoon et~al.}{2007}]{2007ApJ...654L..49S}
Spoon H.~W.~W.,  Marshall J.~A.,  Houck J.~R.,  Elitzur M.,  Hao L.,  Armus L.,
   Brandl B.~R.,    Charmandaris V.,  2007, \apjl, 654, L49

\bibitem[\protect\citeauthoryear{Stasi{\'n}ska, Asari, Fernandes, Gomes,
  Schlickmann, Mateus, Schoenell \& Sodr{\'e}}{Stasi{\'n}ska
  et~al.}{2008}]{Stasinska:2008p4785}
Stasi{\'n}ska G.,  Asari N.~V.,  Fernandes R.~C.,  Gomes J.~M.,  Schlickmann
  M.,  Mateus A.,  Schoenell W.,    Sodr{\'e} L.,  2008, \mnras, 391, L29

\bibitem[\protect\citeauthoryear{Veilleux, Kim \& Sanders}{Veilleux
  et~al.}{2002}]{Veilleux:2002p4924}
Veilleux S.,  Kim D.-C.,    Sanders D.~B.,  2002, \apjs, 143, 315

\bibitem[\protect\citeauthoryear{Wild, Heckman \& Charlot}{Wild
  et~al.}{2010}]{Wild:2010p4207}
Wild V.,  Heckman T.,    Charlot S.,  2010, arXiv, astro-ph.CO

\bibitem[\protect\citeauthoryear{Wild, Heckman, Sonnentrucker, Groves, Armus,
  Schiminovich, Johnson, Martins \& LaMassa}{Wild
  et~al.}{2010}]{Wild:2010p4205}
Wild V.,  Heckman T.,  Sonnentrucker P.,  Groves B.,  Armus L.,  Schiminovich
  D.,  Johnson B.,  Martins L.,    LaMassa S.,  2010, arXiv, astro-ph.CO

\bibitem[\protect\citeauthoryear{Wild, Kauffmann, Heckman, Charlot, Lemson,
  Brinchmann, Reichard \& Pasquali}{Wild et~al.}{2007}]{wild_psb}
Wild V.,  Kauffmann G.,  Heckman T.,  Charlot S.,  Lemson G.,  Brinchmann J.,
  Reichard T.,    Pasquali A.,  2007, \mnras, 381, 543

\bibitem[\protect\citeauthoryear{Wild, Walcher, Johansson, Tresse, Charlot,
  Pollo, F{\`e}vre \& de Ravel}{Wild et~al.}{2009}]{Wild:2009p2609}
Wild V.,  Walcher C.~J.,  Johansson P.~H.,  Tresse L.,  Charlot S.,  Pollo A.,
  F{\`e}vre O.~L.,    de Ravel L.,  2009, \mnras, 395, 144

\bibitem[\protect\citeauthoryear{Yu \& Tremaine}{Yu \&
  Tremaine}{2002}]{Yu:2002p3731}
Yu Q.,  Tremaine S.,  2002, \mnras,
  335, 965

\bibitem[\protect\citeauthoryear{Zheng, Bell, Somerville, Rix, Jahnke,
  Fontanot, Rieke, Schiminovich \& Meisenheimer}{Zheng
  et~al.}{2009}]{Zheng:2009p4373}
Zheng X.~Z.,  Bell E.~F.,  Somerville R.~S.,  Rix H.-W.,  Jahnke K.,  Fontanot
  F.,  Rieke G.~H.,  Schiminovich D.,    Meisenheimer K.,  2009, \apj, 707, 1566

\end{thebibliography}

\appendix

\newpage
\section{Sample information}

\begin{table*}
  \centering
  \caption{\label{tab:psb} \small The \hds\ sample. Columns are: name [1], RA [2], Dec [3],
    SDSS specobjid [4], SDSS-MPA rowindex [5], redshift from SDSS [6]}
\vspace{0.2cm}
  \begin{tabular}{ccccccc} \hline\hline
Name & RA (deg) & Dec (deg) & specobjid & rowindex & z \\\hline
PSB01&181.225&52.395&248485859366535168&284940&0.0632\\
PSB02&169.600&56.036&256085720727289856&296570&0.0680\\
PSB03&173.251&60.275&268189274979958784&315471&0.0644\\
PSB04&167.291&45.857&404988882665340928&491844&0.0636\\
PSB05&136.004&1.458&132797726868897792&105860&0.0535\\
PSB06&120.199&37.729&213582458531086336&232315&0.0416\\
PSB07&339.726&13.194&208235580251701248&223368&0.0628\\
PSB08&128.311&32.583&262560681922920448&306604&0.0558\\
PSB09&239.431&27.465&391758540508758016&472114&0.0316\\
PSB10&139.864&33.791&448335754997792768&536057&0.0193\\
PSB11&158.162&12.178&450306505485320192&539391&0.0329\\
  \end{tabular}
\end{table*}

\begin{table*}
  \centering
  \caption{\label{tab:ul} \small The ULIRG sample, where SDSS spectra
    exist. Columns are: name as given in Armus et al, (2007) [1], RA [2], Dec [3],
    SDSS specobjid [4], SDSS-MPA rowindex [5], redshift from SDSS
    [6].}
\vspace{0.2cm}
  \begin{tabular}{ccccccc} \hline\hline
Name & RA (deg) & Dec (deg) & specobjid & rowindex & z \\\hline
IRAS 08572+3915&135.106&39.065&337714854909444096&394321&0.0579\\
IRAS 12112+0305&183.441&2.811&146028589375029248&126942&0.0730\\
IRAS 15250+3609&231.748&35.977&394574695615168512&476759&0.0553\\
Arp220&233.738&23.504&609061542893191168&755340&0.0181\\
Mrk273&206.176&55.887&372055066565148672&439991&0.0373\\
UGC5101&143.965&61.353&137019791402598400&111495&0.0393\\

  \end{tabular}
\end{table*}

\begin{table*}
  \centering
  \caption{\label{tab:sey} \small The Seyfert sample. Columns are: name
    as given in \citet{2010arXiv1007.0900L}[1], RA [2], Dec [3],
    SDSS specobjid [4], SDSS-MPA rowindex [5], redshift from SDSS
    [6].}
\vspace{0.2cm}
  \begin{tabular}{ccccccc} \hline\hline
Name & RA (deg) & Dec (deg) & specobjid & rowindex & z \\\hline
2MASXJ08244333+2959238&126.180&29.990&339966523504328704&397741&0.0250\\
CGCG064-017&149.812&12.988&491120230488080384&603606&0.0340\\
Mrk0609&51.356&-6.144&129701503752470528&101843&0.0340\\
NCG0291&13.375&-8.768&185153160128495616&190581&0.0190\\
NGC5695&219.342&36.568&389226545811554304&467780&0.0140\\
2MASXJ15522564+2753435&238.107&27.895&391758541012074496&472216&0.0740\\
2MASXJ12384342+0927362&189.681&9.460&347285137836212224&409636&0.0830\\
2MASXJ08035923+2345201&120.997&23.756&356292212926971904&415287&0.0290\\
2MASXJ16164729+3716209&244.197&37.273&297464197648744448&366718&0.1520\\
2MASXJ11570483+5249036&179.270&52.818&248485858699640832&284823&0.0360\\
2MASXJ12183945+4706275&184.664&47.108&408648327781941248&498207&0.0940\\
SBS1133+572&173.954&56.952&369240319515951104&435171&0.0510\\
SBS1609+490&242.716&48.911&175301007746531328&174339&0.0450\\
IC0486&120.087&26.614&261716087635181568&305068&0.0270\\
CGCG242-028&170.755&47.052&405833433537839104&493622&0.0250\\
Mrk1457&176.840&52.450&247922896524017664&283914&0.0490\\
2MASXJ13463217+6423247&206.634&64.390&170234565783191552&166011&0.0240\\
CGCG218-007&200.952&43.301&387537584101785600&464960&0.0270\\
2MASXJ10181928+3722419&154.580&37.378&401892408733728768&486221&0.0490\\
2MASXJ11110693+0228477&167.779&2.480&143495710086529024&122473&0.0350\\
  \end{tabular}
\end{table*}

\section{\hds\ galaxies}\label{app:psb}
While three of the samples studied in this paper have been presented
elsewhere, the \hds\ sample has not. Here we summarise its selection
criteria and the mid-IR observations.

\subsection{Selection}
The galaxies were selected to have:
\begin{itemize}
\item Observed \ha/\hb\ flux ratios greater than 8
\item Strong Balmer absorption lines compared to their 4000\AA\ break
  strength [${\rm PC2}>0.05\times {\rm PC1}+0.4$,
  \citet{wild_psb}]. Although they were not selected on a single
  Balmer absorption line equivalent width, which have considerably
  larger errors than the spectral indices used here, most have
  H$\delta_A>5$\AA\ and $D_n(4000)<1.4$. 
\item Accurately measured emission lines: a signal-to-noise $>3\sigma$
  for all four BPT lines, together with a \hb\ flux $\rm
  >40\times10^{-17} erg/s/cm^2$.
\item A contribution to their emission lines by an AGN, i.e. they lie
  above the empirically derived demarcation line of
  \cite{2003MNRAS.346.1055K} on the BPT diagram.
\item $z<0.07$ so that the SDSS spectrum, taken through a 3\arcsec
  fibre, samples only light coming from the central few kpc of the
  galaxy.
\end{itemize}

\subsection{Mid-IR observations}

IRS observations in both high and low resolution were carried out
using staring mode during program 40330 (PI Timothy Heckman).

We used SPICE to extract the 1D spectra from the low-resolution basic
calibrated data (BCD) pipeline products, using the point-source and
optimal extraction options. The two 1D spectra (i.e. one for each nod)
were combined using a weighted mean. For the high-resolution spectra,
we first identified and interpolated over additional rogue pixels
using a custom IDL routine and the rogue pixel masks suitable for our
campaign numbers (41-46). This procedure was repeated for the
dedicated off-source background observations. All 3 science exposures
were combined using a simple arithmetic mean, resulting in two 2D
spectra. These were sky-subtracted and the 1D spectra were extracted
using SPICE with the point-source and optimal extraction options. For
each nod, the edges of each order were inspected and trimmed and the
orders combined using a weighted mean where wavelengths were covered
by more than one order. The 1D spectra of each nod were subsequently
combined using a weighted mean.

\end{document}